\documentclass[9pt,twocolumn,twosides]{pnas-new}
\usepackage{graphics,color,graphicx,amsmath}
\usepackage{xr}
\usepackage{rotating}[counterclockwise]
\usepackage{authblk}
\usepackage{url}
\usepackage{hyperref}
\usepackage{multirow}
\usepackage{adjustbox}

\newcommand{\Dt}[0]{a}
\newcommand{\Dx}[0]{b}

\bibstyle{unsrt}
\templatetype{pnasresearcharticle} 

\significancestatement{While infamous wars like World War I are seemingly coherent, they are actually composed of a variety of events that have been clustered together by choice. How they should cluster is complicated because events may be instantaneous like an explosion, but may result from years of building tension. Geographically, violence may occur on a single street, but is it part of a revolution? The range of temporal and spatial scales poses a challenge for deciding if events belong together. We address this problem by developing a statistical approach for uncovering chains of related conflict events that accounts for scale. We discover that coherent chains of conflict events predominate in the timescale of a week to months and around the distance separating urban areas of populations with 100{,}000 to a million people. Within this middle scale, our approach connects events in a way that highlights plausible and potentially hidden causal mechanism. Using Nigeria, Somalia, and Sierra Leone as examples, we develop a way of visualizing the certainty and range of connections. Our method helps uncover and exclude causal links in conflict spread and can be applied to other spreading social phenomena.}

\title{Discovering the mesoscale for chains of conflict}
\author{Niraj Kushwaha}
\author{Edward D.~Lee}
\affil{Complexity Science Hub, Josefst{\ae}dter Strasse 39, Vienna, Austria}

\dates{This manuscript was compiled on \today}
\leadauthor{Kushwaha}

\authorcontributions{E.D.L.~designed the research. N.K.~and E.D.L.~wrote the code, did the analysis, and drafted the manuscript.}

\authordeclaration{The authors declare no competing interests.}
\correspondingauthor{\textsuperscript{2}To whom correspondence should be addressed. E-mail: edlee@csh.ac.at}


\date{\today}

\begin{abstract}
Conflicts, like many social processes, are related events that span multiple scales in time, from the instantaneous to multi-year developments, and in space, from one neighborhood to continents. Yet, there is little systematic work on connecting the multiple scales, formal treatment of causality between events, and measures of uncertainty for how events are related to one another. We develop a method for extracting related chains of events that addresses these limitations with armed conflict. Our method explicitly accounts for an adjustable spatial and temporal scale of interaction for clustering individual events from a detailed data set, the Armed Conflict Event \& Location Data Project. With it, we discover a {\it mesoscale} ranging from a week to a few months and from tens to a few hundred kilometers, where long-range correlations and nontrivial dynamics relating conflict events emerge. Importantly, clusters in the mesoscale, while extracted only from conflict statistics, are identifiable with causal mechanism cited in field studies. We leverage our technique to identify zones of causal interaction around conflict hotspots that naturally incorporate uncertainties. Thus, we show how a systematic, data-driven procedure extracts social objects for study, providing a scope for scrutinizing and predicting conflict amongst other processes.
\end{abstract}

\begin{document}
\maketitle

\thispagestyle{firststyle}
\ifthenelse{\boolean{shortarticle}}{\ifthenelse{\boolean{singlecolumn}}{\abscontentformatted}{\abscontent}}{}

\dropcap{H}istorically, the study of armed conflict has focused on pre-defined aggregates like skirmishes, battles, and wars, where individual acts of violence have been integrated into a coherent whole by experts \cite{palmer2013history,laforeLongFuse1997,richardsonVariationFrequency1948}. Yet, such a procedure is difficult to replicate systematically across time periods and regions because it is fundamentally qualitative. More recently, sensitivity to the underlying assumptions in the definition of conflict has inspired the creation of ``disaggregated'' data sets, where the atomic units, or events, are delimited by a location, time, and other distinguishing characteristics  \cite{raleighIntroducingACLED2010,balcellsViolenceCivilians2021,balcellsViolenceCivilians2021}. Naturally, disaggregated data introduce the complementary difficulty of clustering events into meaningful conflict aggregates \cite{balcellsViolenceCivilians2021,buhaugAccountingScale2005}. Generally, heuristics are used to group events together using properties like involved actors \cite{dowdCulturalReligious2015}, geographical boundaries \cite{raleighPoliticalHierarchies2014}, administrative boundaries \cite{machClimateRisk2019}, or ethnic divisions \cite{michalopoulosLongRunEffects2016}. While these approaches are helpful for building intuition, they are not considered systematic in the conflict literature \cite{raleighIntroducingACLED2010, kikutaNewGeography2022,raleighPoliticalHierarchies2014,hegrePovertyCivil2009}, the groupings are fixed and rigid, and they can be sensitive to the way that the data are labeled, which is subject to purposeful or inadvertent errors. In short, there is a need for a systematic procedure for dealing with scale that goes beyond qualitative treatments \cite{hegrePovertyCivil2009}, a quantitative framework for extracting causal relationships, and consequently a provision for uncertainty in the inferred relationships between conflict events. Such a technique would be useful not only for the study of political violence but more generally for other social processes that spread across time and space.

We demonstrate here a systematic procedure that addresses these limitations by uncovering causal patterns from conflict statistics. Our approach is inspired by fundamental advances in physics and biophysics relating to the analysis of multiple scales in cascades such as the propagation of stress in collapsing materials and neural activity in the brain \cite{sethnaCracklingNoise2001,jensenSelforganizedCriticality1998,beggsNeuronalAvalanches2003, williams-garciaUnveilingCausal2017,meshulamCoarseGraining2019}. 
Our approach is robust to errors because it relies only on information about the presence or absence of conflict, introduces a distance-dependent measure of causal influence incorporating uncertainties, and allows analyses to move systematically between spatial and temporal scales.

\begin{figure*}\centering
	\includegraphics[width=.8\linewidth,clip=true,trim=0 80 0 100]{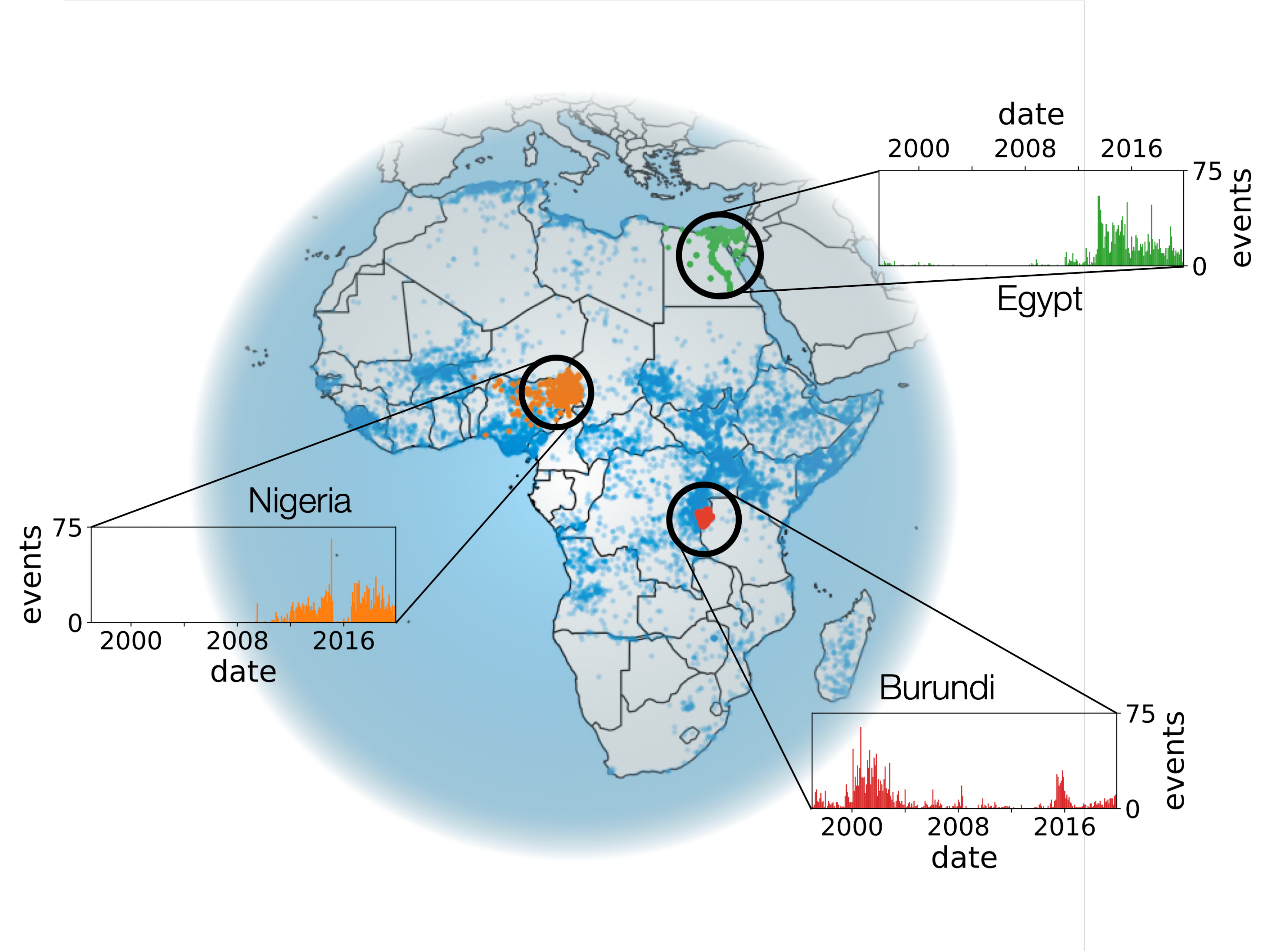}
	\caption{Spatial and temporal distribution of conflict events included in the Armed Conflict Location \& Event Data Project (ACLED) from 1997 through 2019 in Africa. Each point is a location at which conflict has been reported. For the three regions of northeastern Nigeria, Egypt, and Burundi, we show the monthly incidence of reported conflict events.}\label{fig:data}
\end{figure*}

We focus on the Armed Conflict Location \& Event Data Project (ACLED), which provides an extensive, publicly available, and disaggregated dataset on worldwide conflict \cite{raleighIntroducingACLED2010}. Each conflict event noted in the database occurs in a particular time and place between a set of actors constituting a point of activity as plotted on the map in Figure~\ref{fig:data}. By summing over the points of activity in a particular region, we are also able to track levels of conflict over time as in the insets. Information about conflict events is collected from news and local sources, and the database details for each event alleged actors, fatalities, location, date, and precision of the provided data. Importantly, the events between armed groups are labeled as ``battles,'' allowing us to focus on them. 
Amongst the battles, we analyze conflict in Africa because it is there where we have the longest observational period (1997-2019) and a large contiguous landmass compared to other regions. As a result, the data set provides a high-resolution perspective on the atomic units of conflict that we can use to determine how events should be joined together. 

As an example of what one would like to analyze, we highlight attacks labeled ``Boko Haram'' in northeastern Nigeria in orange in Figure~\ref{fig:data}. Neither grouping conflicts by major militant group Boko Haram nor by country boundaries captures their relationship to surrounding areas; for example, field research has indicated that the group drives conflict in western Chad by forcing herders to migrate south and east \cite{georgeExplainingTranshumancerelated2022}. At a wider scale, violence perpetrated by Boko Haram impacts conflict prevalence elsewhere in Nigeria, if indirectly, because such events tend to sap government resources, erode government legitimacy, and cause economic damage \cite{weinsteinRebellion2007,corralFragilityConflict2020,innocentCostBoko2012}, connections that are not apparent from this grouping. Inspecting the spatial distributions in Figure~\ref{fig:data}, we see that conflict events tend to cluster with one another in time and space \cite{leeScalingTheory2020}. This suggests that the way that conflict may drive more conflict would be detectable in local statistical patterns of activity.

\begin{figure*}[t]\centering
    \includegraphics[width=\linewidth]{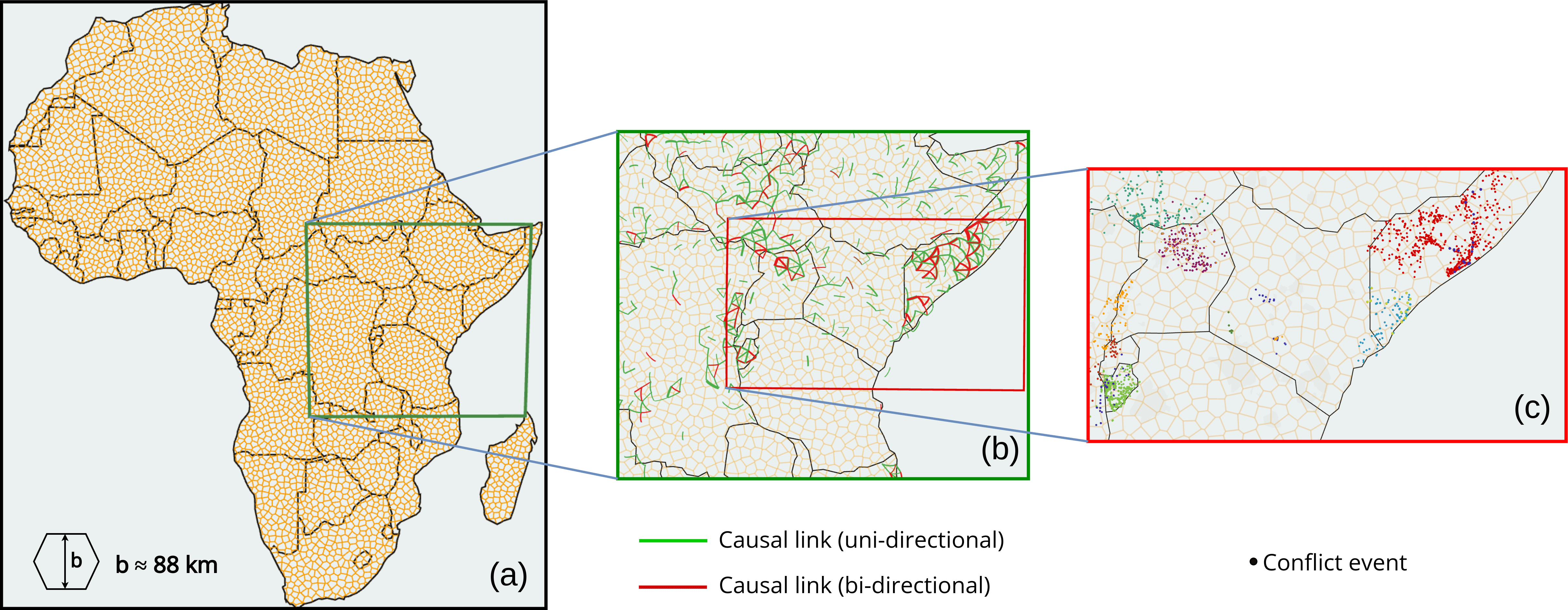}
    \caption{Illustration of conflict avalanche generation (see Appendix~\ref{Sec: Avalanche generation} for detailed algorithm). (a) We first set the spatial and temporal separation scales. (b) We then infer causal structure by calculating directed transfer entropy for pairs of neighboring spatial bins. An example of the causal network is shown for temporal scale $a=64$ days and spatial scale $b \approx 88$ km. Links are directed in nature but the arrows are not shown here for simplicity. (c) Conflict avalanches are sequences of conflict events (points on map) that are connected through the causal network. Different colors correspond to different avalanches.}\label{fig:method}
\end{figure*}

We leverage local conflict patterns to extract a causal geographic web identifying paths through which conflicts might affect each other. Building on previous work, we set spatial and temporal separation scales, $\Dx$ and $\Dt$, grouping together conflict events that fall within the specified distance of one another \cite{leeScalingTheory2020}. This is akin to establishing a minimal resolution in our viewing lens, or a scale on a spatial kernel or temporal memory, such that events that are closer together than this distance cannot be distinguished from one another. Here, we perform such a discretization using temporal bins of duration $\Dt$ and pseudorandom Voronoi cells with typical radius $\Dx$ to avoid artifacts from regular lattices. We show examples of the cells in Figure~\ref{fig:method} (more details on the algorithm in Appendix~\ref{Sec: Avalanche generation}). Our procedure allows us to titrate the coarseness of our resolution with precisely defined scales, at the smallest scales grouping only local conflict events together and at the largest allowing for the possibility that conflict events belong together across continental distances and years.

\begin{figure*}\centering
	\includegraphics[width=.9\linewidth]{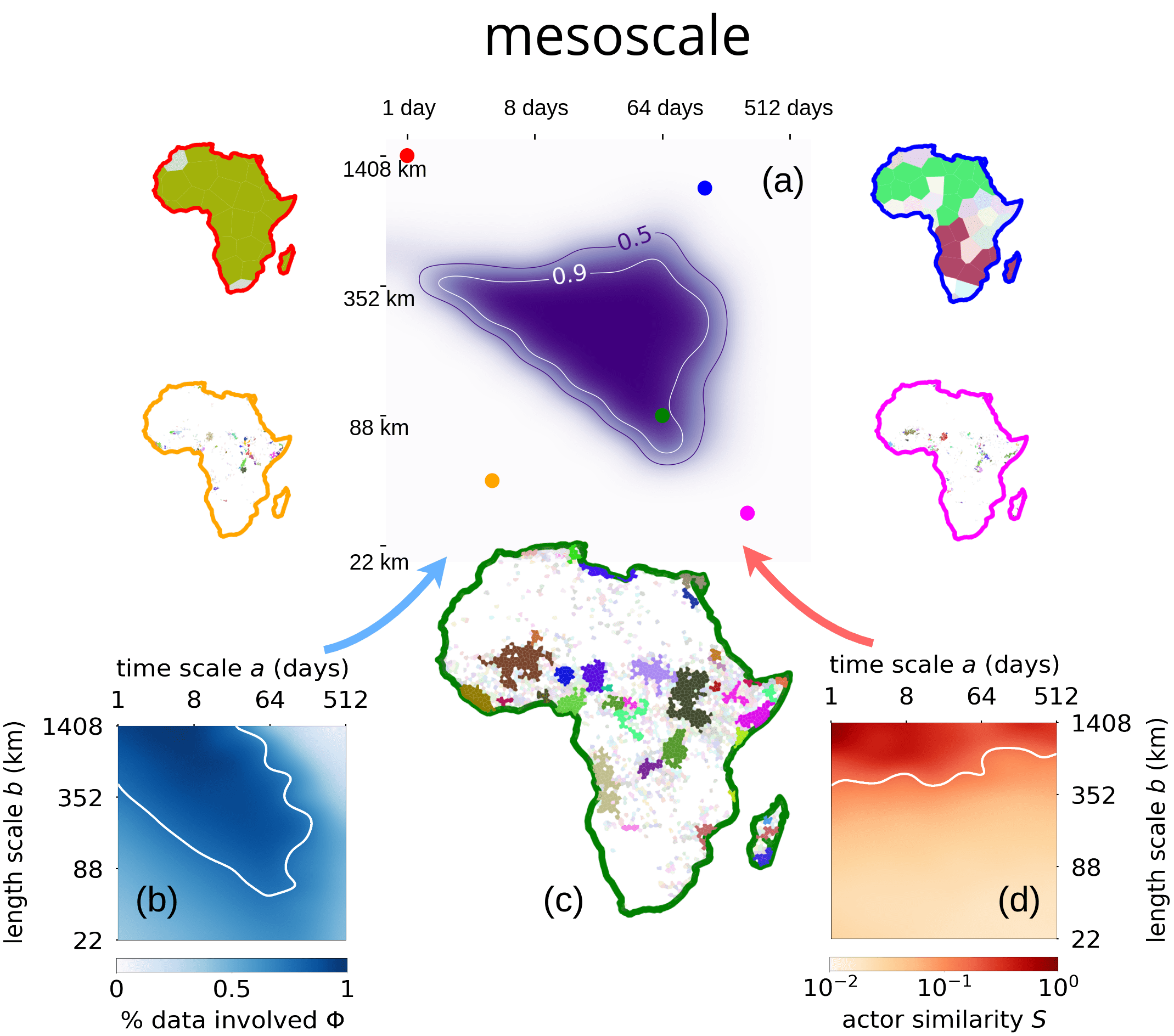}
	\caption{Mesoscale for armed conflict. (a) Mesoscale is identified using scales in which conflict avalanches (b) contain more than 3/4 of the data $\Phi\geq3/4$ and (d) actor similarity is less than the midpoint of actor similarity score $S\lesssim 0.132\pm0.002$, with standard deviation given over Voronoi tessellations. (c) Example of conflict zones in the mesoscale, $a=64$\,days, $b\approx88$\,km. The mesoscale in (a) is obtained using an overlap of 20 different realizations of Voronoi tessellations (For details see Appendix~\ref{sec: interpolating}. For a few examples, we display the resulting conflict zones as indicated by the markers in panel a, where color corresponds to the outline of the respective African map. Colors inside each conflict zone map corresponds to different zones. Conflict zones which span less than five Voronoi cells are shown in faint colors.}\label{fig:mesoscale}
\end{figure*}

\begin{figure}[t!]\centering
    \includegraphics[width=.8\linewidth]{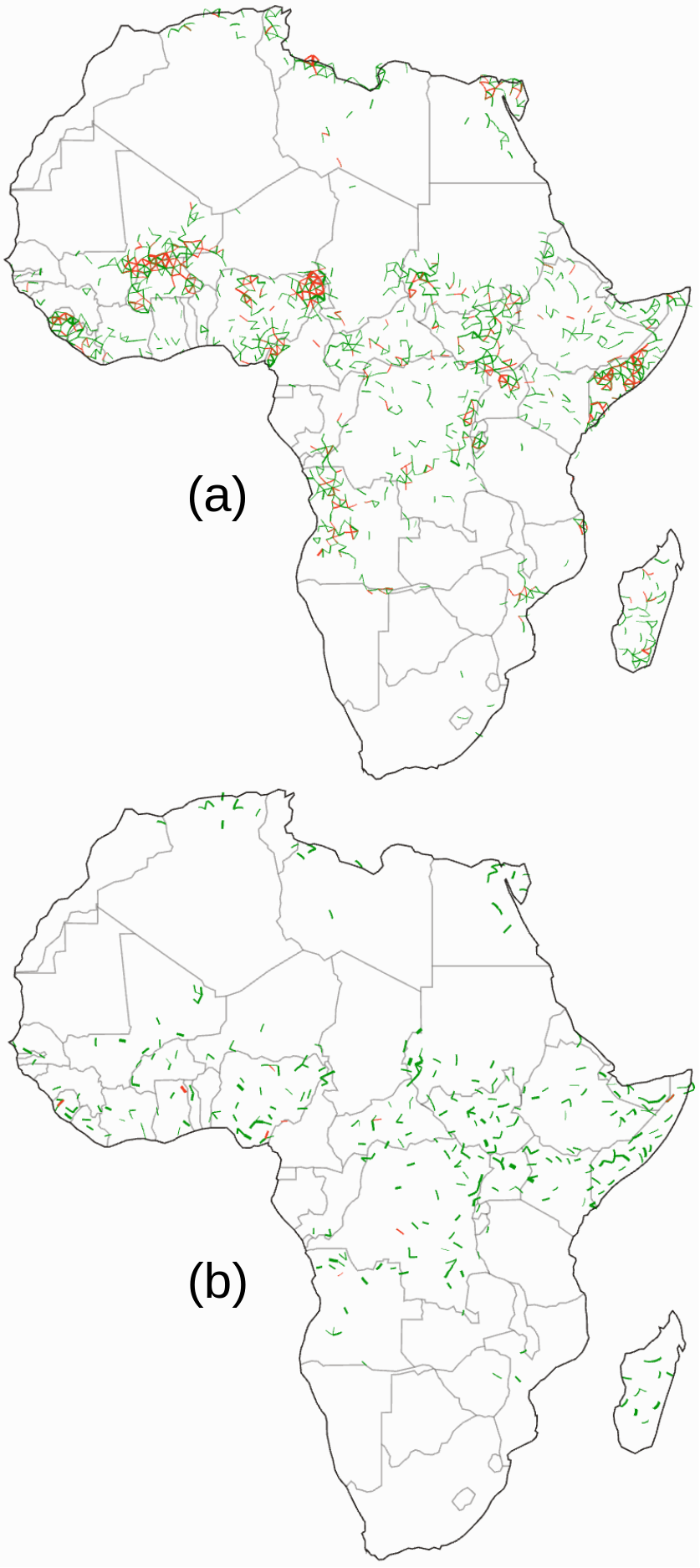}
    \caption{Causal network through which conflict avalanches propagate. (a) Statistically significant causal edges between adjacent Voronoi cells using transfer entropy ($\Dt=64$\,days, $\Dx\approx88$\,km). Directed nature of graph not shown. Edges shown in green have a causal edge in one direction only, red in both directions. (b) Causal network from time-shuffled null model is fragmented.}
\label{fig:network}
\end{figure}

\begin{figure*}[t]\centering
	\includegraphics[width=\linewidth]{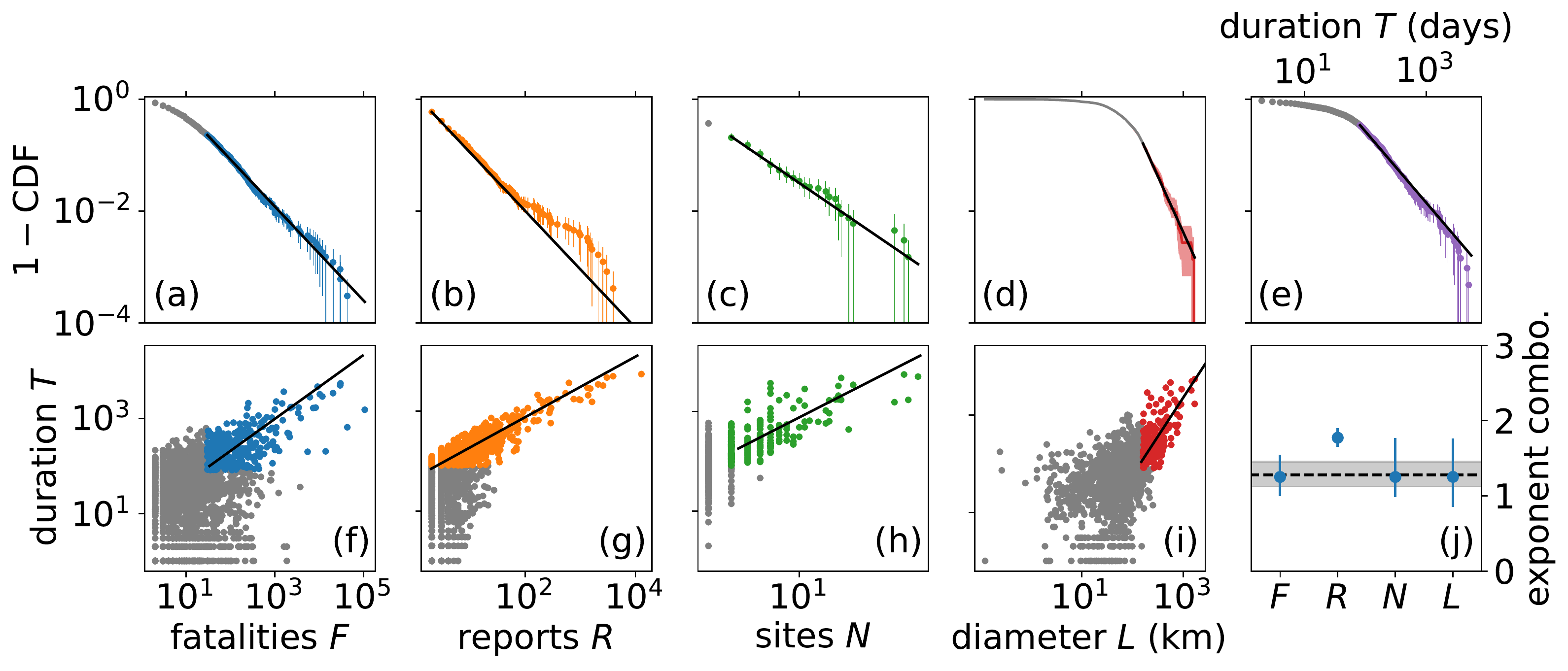}
	\caption{Scaling in the mesoscale for $\Dx=176$\,km, $\Dt=64$\,days. (a-e) Distributions of avalanche properties have exponents $\tau=1.8\pm0.1$ ($p=0.73$), $\tau'=2.07\pm 0.03$ ($p=0.002$), $\mu=2.7\pm0.4$ ($p=0.56$), $\nu=3.0\pm0.4$ ($p>0.99$), $\alpha=2.3\pm0.1$ ($p=0.075$). A $p$-value of greater than 0.1 is significant, which indicates that some of the fits are only approximately power laws \cite{clausetPowerLawDistributions2009}. Points below the lower cutoff are gray. Exponent error bars represent one standard deviation over $10^3$ bootstrapped samples. Shown error bars in plan correspond to 95\% confidence intervals over the same bootstrapped samples. (f-i) Duration vs.~conflict measures, or dynamical scaling. Points below the respective lower cutoffs in the power law distributions are not fit and are shown in gray. (j) Predicted exponent relations relating exponent for duration distribution $\alpha$ (shaded region) vs.~exponent combination for remaining variables (markers) align for all except reports $R$, which deviates from a power law distribution.}\label{fig:scaling}
\end{figure*}

For a given scale, we determine whether or not a particular Voronoi cell $x_t$ had any conflict at some moment in time bin indexed $t$ in which case $x_t=1$; otherwise, $x_t=0$. This presents a binary time series, where the pattern of activity reveals interaction between conflict in time and space. As a pragmatic hypothesis that limits the range of possible causal interactions (and thus many false positive and negatives), we start with the assumption of local causality, or that conflicts in one zone $x$ are potentially affected only by neighboring zones $y$ defined as cells that touch. The simplest case is if no neighboring influence exists such that conflict at one site influences itself in the future. In other words, we would expect that if the presence of conflict in the future of site $x$, denoted as $x_{t+1}$, depended on the past $x_t$, then the relationship between the joint probabilities $q(x_t,x_{t+1})$ would not factorize into the marginals, or $q(x_t,x_{t+1})\neq q(x_t)q(x_{t+1})$. This difference is given by the mutual information between past and future \cite{coverElementsInformation2006}
\begin{align}
	I[X_t; X_{t+1}] &= \sum_{\substack{x_t\in \{0,1\}\\x_{t+1}\in\{0,1\}}} q(x_t, x_{t+1}) \log\left( \frac{q(x_t,x_{t+1)})}{q(x_t)q(x_{t+1})} \right).\label{eq:I}
\end{align}
On the other hand, it could have been the case that the neighborhood played a role such that having information about a neighbor at the present $y_t$ helps predict what happens in the future $x_{t+1}$. This is exactly the quantity described by the transfer entropy \cite{schreiberMeasuringInformation2000}, which tells us if knowing about neighboring cell $y_t$ conveys any further information about $x_{t+1}$ beyond what was already given by $x_t$,
\begin{align}
	T[X;Y] &= \sum_{x_t,x_{t+1},y_t} q(x_t, x_{t+1}, y_{t}) \log \left(\frac{q(x_{t+1}|x_{t},y_{t})}{q(x_{t+1}|x_{t})}\right).\label{eq:te}
\end{align}
Transfer entropy is zero when $q(x_{t+1}|x_t,y_t) = q(x_{t+1}|x_t)$. Unlike Granger causality, transfer entropy is a nonlinear and general measure of statistical dependence and the two are equivalent only for Gaussian variables \cite{barnettGrangerCausality2009}. Finally, we must worry about the fact that we have a finite time series on which to calculate Eqs~\ref{eq:I} and \ref{eq:te}. To take this into account, we test the significance of the mutual information and transfer entropy by asserting that they are only significantly different from zero when at least a fraction $1-p$ of time-shuffled values are smaller than the measured value \cite{papanaReducingBias2011}. As we adjust the significance threshold $p$, we go from allowing any pair of proximate conflict events to be connected, $p = 1$, to sparse connections, $p=0$. Here, we only take edges as causal when they are significant with the cutoff $p\leq1/20$. This dramatically limits the number of causal connections, indicating that only a sparse set of the possible edges between cells demonstrate recognizable causal signal. 

The causal signals that we find tend to cluster in geographic regions including the Sahel from multiple ongoing conflicts \cite{wehreyPerilousDesert2013,raleighPoliticalMarginalization2010}, northeastern Nigeria from Boko Haram \cite{perousedemontclosBokoHaram2014}, Nigeria and Cameroon from Ambazonian Separatist \cite{ekahAnglophoneCrisis2019}, Northern Africa from civil war \cite{kimMiddleEast2020}, the Horn of Africa from state failure and insecurity \cite{menkhausSomaliaHorn2011}, the Darfur region from genocide and ethnic hostilities \cite{bassilPostcolonialState2015}, Angola and Congo from civil war \cite{guimaraesOriginsAngolan2016}, and Madagascar from the Dahalo Militia \cite{hemezCoreeSud2017} as we show in Figure~\ref{fig:network}a. In contrast, a null model where we have time shuffled all the events in each Voronoi cell leads to a dispersed and fragmented causal network as we show in Figure~\ref{fig:network}b. That the highly connected regions represent recognizable conflict ``hotspots'' confirms the power of our systematic procedure only accounting for statistical signatures of causality in observed conflict events.

\begin{figure*}
	\includegraphics[width=\linewidth]{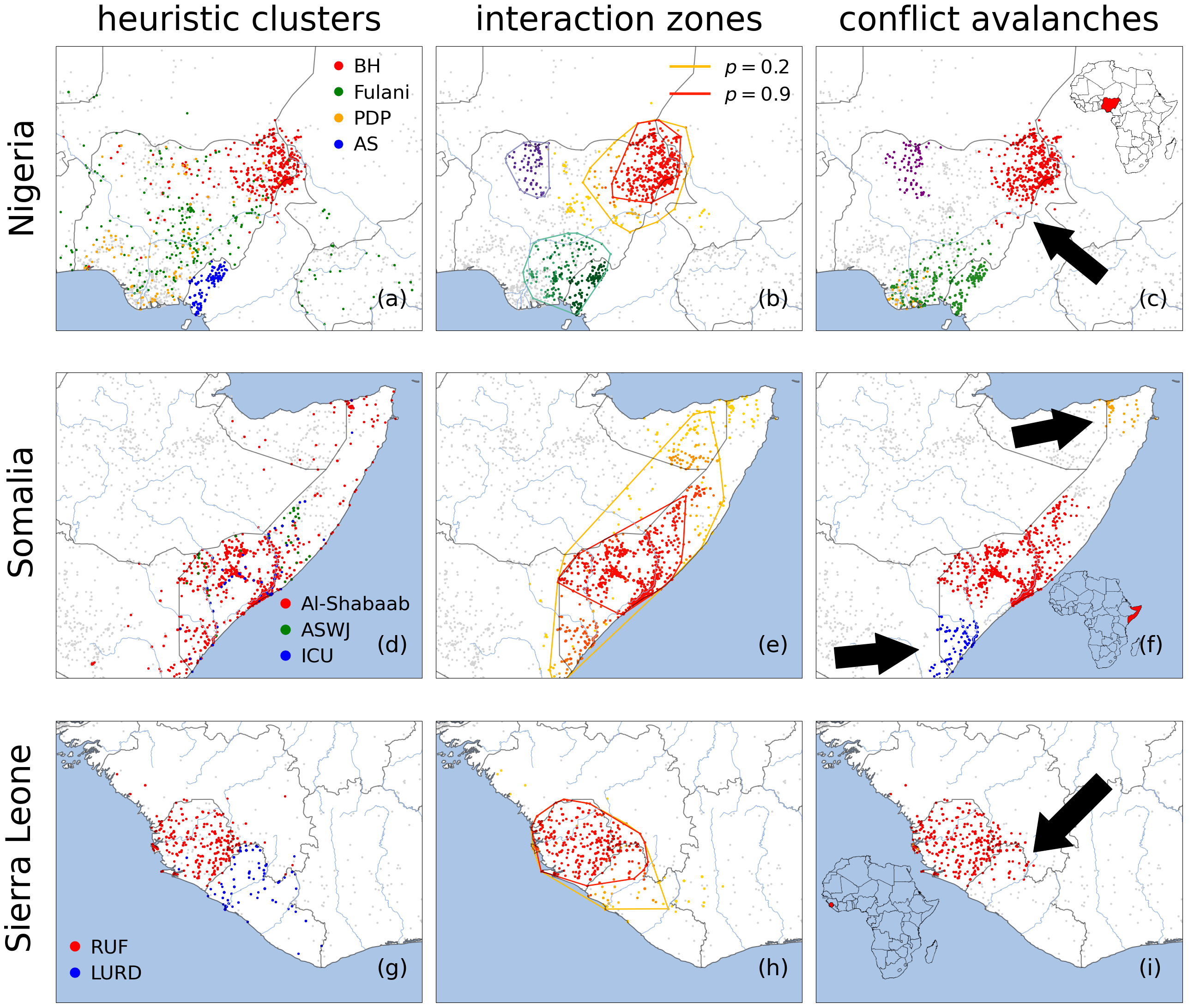}
	\caption{Heuristic conflict clusters vs.~systematic conflict avalanches. Gray points represent all the conflict events that are not part of the shown conflict clusters and avalanches. Conflict in (a-c) Nigeria with $\Dx\approx88$\,km, $\Dt=64$\,days, (d-f) Somalia with $\Dx\approx88$\,km, $\Dt=64$\,days, and (g-i) Sierra Leone with $\Dx\approx66$\,km, $\Dt=64$\,days. Conflicts identified by (a,d,g) actor names, (b,e,h) conflict interaction incorporating probability of an event being part of the highlighted conflict avalanche, (c,f,i) biggest conflict avalanches (in terms of geographic spread) in the region. Arrows point to conflict events that are associated with (c) Fulani Militia but are grouped with the Boko Haram cluster in panel a, (f) Al-Shabaab attacks which are separate from the central Al-Shabaab cluster in red, and (h) LURD events combined with RUF. (b) Purple and green convex hulls correspond to $p=0.5$. (BH$=$Boko Haram, PDP$=$People's Democratic Party, AS$=$Ambazonian Separatists, ASWJ$=$Ahlu Sunna Waljama'a, ICU$=$Islamic Courts Union, RUF$=$Revolutionary United Front, LURD$=$Liberians United for Reconciliation and Democracy)}\label{fig:regions}
\end{figure*}

The resulting causal network gives us a way of tracing causal chains of conflict events. We connect all events to one another that have occurred together in the same spatiotemporal bin and that have occurred in any adjacent spatial bin at a sequential time to which there is an outgoing causal edge. From such a procedure we obtain clusters that cascade over time, or {\it conflict avalanches} (see movie in reference \citenum{kushwahaConflictAvalanche2022}), across a wide range of scales as we vary $\Dx$ between $10$\,km to $10^3$\,km (Africa is about $10^4$\,km wide) and $a$ from $1$ day to $10^3$ days.
As one picture of conflict avalanche extent, we color the geographic regions that a conflict avalanche covers in totality, joining together regions when avalanches intersect with one another to define {\it conflict zones}. These are the colored regions in Figure~\ref{fig:mesoscale}c. We find that at the largest separation scales nearly all of Africa is lumped together into a single large conflict zone (top left map in panel a), whereas at the smallest scales Africa fragments into small disparate zones (bottom right map panel in panel a). Only in between the extremes do we find conflict avalanches covering a wide range of scales, displaying scaling statistics, and whose spatial extents are qualitatively recognizable as in Figure~\ref{fig:mesoscale}c. This suggests the existence of some mesoscale at which conflict avalanches correspond to meaningful narratives of cause and effect.

To make this intuition more concrete, we propose a simple first-principles argument for isolating a mesoscale. First, we stipulate that most conflict events should belong in a conflict avalanche; otherwise, conflict avalanches are not a useful representation of the data. Since majority could mean anywhere from half to all of the data, we choose the midpoint at $\Phi=3/4$, or that at least that fraction $\Phi$ of the data must belong in a conflict avalanche. As we show in Figure~\ref{fig:mesoscale}b, this threshold (the white line) delineates a region of scales in the upper left portion of the full space.

Second, we remark that the largest conflict avalanches tend to group disparate actors together even though conflict actors tend to be geographically localized. We quantify this intuitive criterion by defining an actor similarity score $S$, which gauges how similar sets of actors are between adjacent conflict zones. A standard metric would be to consider the normalized overlap between sets of actors between zones, but this fails to account for the possibility that some actors overwhelmingly dominate the set of observed events, whereas others may only appear once or twice. To account for this imbalance, we compute a weighted overlap that accounts for the fraction of events in which actors are involved (see Appendix~\ref{sec:actor similarity}). Our similarity score is the average over all such pairwise comparisons including self comparisons such that $S=0$ when none of the conflict zones have overlapping actors, and it saturates at $S=1$ when all conflict zones have the same actor distribution amongst events. Therefore, similarity provides a normalized measure that accounts for how homogeneous or heterogeneous conflict zones are from one another as we show in Figure~\ref{fig:mesoscale}d.

Since continent-encompassing conflict zones show maximal diversity, uniting nearly all actors into a single large conflict, whereas fragmented zones fail to connect events perpetrated by the same actor, we choose the logarithmic midpoint of similarity $S$, which is about $S\approx0.132$.\footnote{The logarithm accounts for the fact that similarity $S$ is heavily skewed towards 0, when we have many avalanches and most pairwise comparison consist of unrelated regions.} In agreement with the observation that actors are mostly geographically localized, we show in Figure~\ref{fig:mesoscale}d that this threshold cuts almost horizontally across at a fixed value of $\Dx\approx350\,$km. Putting the two thresholds together, we obtain in the intersection a {\it mesoscale} that we highlight in Figure~\ref{fig:mesoscale}a, which denotes the region, derived from first-principles, where we anticipate nontrivial examples of conflict to be located.

The boundaries of the mesoscale represent a tradeoff between the spatial and temporal scales of analysis. Its shape indicates that at sufficiently short temporal scales of a few days to a week only a very limited range of geographic scales reveal identifiable and meaningful causal patterns. As we increase the temporal scale to about a month to a few months, however, a much wider window of spatial scales display widespread causal spatial dynamics, suggesting a more fruitful region of study as opposed to other timescales. All together, the spatial scales of interest are limited to between tens to a few hundred kilometers. This indicates that for practical purposes analysis of conflict spread is limited to a range of scales that span $b\approx 60$\,km to $b\approx 400$\,km. To get a sense of what such extracted scales represent, we calculate the typical distance between neighboring populations using ``urban agglomerations,'' a generalized definition of a city as defined in the Africapolis data set \cite{sahelandwestafricaclubAfricapolis2022}, and we find the nearest neighbors for the smallest agglomerations are about $\sim 20$\,km in 2015, which is far below our lower cutoff (see Figure~\ref{fig:africapolis}). Only once we consider agglomerations with at least $10^5$ people do we find that the typical distance coincides with the minimum of our mesoscale at $b\approx60$\,km. The maximum of the mesoscale at $b\approx400$\,km aligns with the distance between pairs of large cities, or at least $10^6$ people. Thus, our method of extracting a mesoscale suggests that large population centers are what typically mediate conflict spread.

A notable feature of the mesoscale is that the conflict avalanches display a range of temporal and structural scales, or non-Gaussian statistics. Motivated by previous work \cite{leeScalingTheory2020}, we compute power law fits to approximate the distributions of the avalanche properties for each combination of separation scales $\Dx$ and $\Dt$: size in terms of fatalities and reports, geographic extent in terms of area and diameter, and duration, examples of which are shown in Figure~\ref{fig:scaling}a-e. For any of these properties $X$, a power law distribution is of the form $P(X) \sim X^{-\alpha}$ with positive exponent $\alpha$ above some lower cutoff $X\geq X_{\rm min}$. We find the fit parameters using a standard procedure involving maximum likelihood for the exponent and a comparison of the Kolmogorov-Smirnov statistic to choose the best lower cutoff \cite{clausetPowerLawDistributions2009}. We find that the majority of scales included in the mesoscale are consistent with displaying power law tails and that the power law is always a better fit than a reasonable alternative model, the lognormal, by the likelihood ratio beyond the lower cutoff (see Appendix~\ref{sec: other distributions} for more details). The power-law tails in the mesoscale indicate that beyond some minimal size conflict avalanches display multiple relevant scales for dynamics and size. 

Just as length and volume are related in a fixed way for physical objects, conflict properties represent different dimensions of the same conflict avalanches and thus the distributions must be connected to one another. We propose to relate them by noting that we typically expect longer conflicts to become larger. This can be expressed as a dynamical scaling hypothesis for fatalities $F$ (as an example) with duration $T$, or that $F\sim T^{d_F/z}$ for a positive scaling exponent $d_F/z$, where $d_F$ is the fractal dimension and $z$ the dynamical exponent. Then, by using the transformation $P(F)dF=P(T)dT$ with the scaling hypothesis, we obtain the exponent relation $\tau-1 = d_F(\alpha-1)/z$, for the power-law model distributions $P(F)\sim F^{-\tau}$ and $P(T)\sim T^{-\alpha}$. We show one example of tests of the exponent relations in Figure~\ref{fig:scaling}j, and they are usually satisfied across the pseudorandom Voronoi tilings for fatalities, sites, and diameter but not for reports (Figure~\ref{fig: dyn_relations}). The violation is unsurprising in that the distribution of reports often displays a substantial hump in the tail that deviates from the power law as in Figure~\ref{fig:scaling}b. This is an indication of yet unexplained mediating variables or processes missing in the dimensional analysis, but these other exponent relations indicate that most aspects of conflict avalanches conform approximately to a low-dimensional theory at sufficiently large scales \cite{richardsonVariationFrequency1948,cedermanModelingSize2003,leeScalingTheory2020}.

Since the mesoscale was extracted from statistical patterns, we check if conflict avalanches highlight causal mechanisms identified in the conflict literature. As we show in Figure~\ref{fig:regions}, identifying conflict clusters by the names of involved actors in Nigeria leads to four major, spatially overlapping conflict clusters for Boko Haram (red), Fulani militia (green), the People's Democratic Party (orange), and Ambazonian separatists (blue). For a choice of separation scales that is comparable, example conflict avalanches in Figure~\ref{fig:regions}c group events differently. Some Fulani militia attacks south of the red Boko Haram cluster form part of the latter rather than a separate group as in panel a. Our clustering is supported by field studies, where it has been pointed out that clashes between Boko Haram and local herders drive the latter further from their normal ranges in northeastern Nigerian leading to conflict between herders and farmers \cite{georgeExplainingTranshumancerelated2022}. Furthermore, we identify the events associated with the conflict at the border of Nigeria and Cameroon as green in panel c, which are generally unrelated to the purple events in the northwest. Taken together, this is consistent with the ``Triangle of Terror'' in Nigeria \cite{omitolaTriangleTerror2021}: Boko Haram (red), Fulani Militia and Anglophone crisis (green), and banditry prevalent in the Zamfara region (purple) \cite{belloFarmersHerdsmen2021}. This particular example confirms that we are able to extract clusters that qualitatively correspond to but also could enhance studied conflict groupings.

As a more systematic look, we can look across the many instances of conflict avalanches that are produced from our algorithm, the exact details of which vary with the randomness in the Voronoi tessellation. By averaging over the different tilings, we measure the strength of the causal connection, or the probability $p$ that any two events are joined into the same conflict avalanche. After calculating this probability for the central events in the Boko Haram avalanche (which always appear together), we draw the convex hulls containing the outermost points for fixed values of $p$ in Figure~\ref{fig:regions}b. These are regions of causal interaction that confirm the core of Boko Haram insurgency that is always grouped together as indicated in red. Further out, the regions reveal a substantially stronger relationship between the core and the events involving the Fulani militia. For comparison, we also show the $p=1/2$ contours for Zamfaran banditry and Ambazonian rebels. Importantly, repeating this exercise across other examples reveals that causal interaction is not simply a function of geographic distance, but events in the same location can trace distinct causal origins. Remarkably, our approach relying only on statistical measures of causality corresponds to causal mechanisms hypothesized in field and conflict studies, suggesting that this provides a powerful scope for identifying hidden interactions. 

For two other examples, we inspect conflict in Somalia and Sierra Leone. In the case of Somalia, we show that different scales reveal underlying structure in the local and regional components of the different subgroups of the Al-Shabaab insurgency. As an example, the bottom arrow in Figure~\ref{fig:regions}e points to conflict events associated with Al-Shabaab which are not clustered with the northern (red) core. Indeed, the former were not part of the initial insurgency and instead caused by conflict with Kenya (blue) when Kenya invaded southern Somalia to ``flush-out'' Al-Shabaab \cite{andersonKenyaWar2015}. Similarly, the top arrow points at a local chain of violence around Bosaso due to the presence of Al-Shabaab and its support groups in the area. This is reflected in the regions of causal interaction that reveal a core of high confidence $p\geq0.9$ along the Shebelle river in the south, which is only weakly linked with events in the north. In Sierra Leone, our clustering procedure strongly suggests ($p\geq0.9$) that the events perpetrated by the Revolutionary United Front (RUF) are related to those in Liberia by Liberians United for Reconciliation and Democracy (LURD). 
Such a relationship is purported by court allegations that Sierra Leone's government helped in the formation of LURD by training fighters in Guinea \cite{sesaySierraLeone2010} and further substantiated by RUF fighters joining LURD \cite{themnerLeapFaith2013}. Another mechanism is alleged fights between RUF supporters and LURD opponents of warlord Charles Taylor in Liberia's Lofa county \cite{CourtHears2021}. In contrast, events in C\^ote D'Ivoire are highly unlikely to be related with $p\leq0.1$ as the dearth of literature on the relation between the two conflicts suggests. Thus, our approach provides a systematic way of measuring area of causal interaction across local to regional scales with a natural measure of uncertainty to mine or disprove causal relationships.



\section*{Discussion}
All conflicts have multiple narrative scales, which can range from the detailed role of the individual (the assassination of Archduke Franz Ferdinand instigating World War I) to geopolitics (a secret alliance network consequently implicating many nation-states \cite{laforeLongFuse1997}) and even further out to societal epochs across civilizational timescales \cite{turchinLinkingMicro2017,fergusonDoomPolitics2021}. Other narratives range from the role of ideologies, personalities, economic incentives, organizational resources and structure, etc.~(see references cited in \cite{weinsteinRebellion2007,raleighPoliticalHierarchies2014}). Each narrative implicitly assumes a relevant range of scales over which to draw a causal relationship between intervening events. And the quantitative evidence confirms that multiple scales matter: many conflict patterns beyond a small size are scale-free, which means that no single scale holds a privileged perspective \cite{johnsonSimpleMathematical2013, leeScalingTheory2020, spagatFundamentalPatterns2018,picoliUniversalBursty2015,clausetTrendsFluctuations2018}. This points to a fundamental challenge in the study of armed conflict, which is that conflict consists of many events occurring at multiple, overlapping spatial and temporal scales \cite{buhaugAccountingScale2005,balcellsViolenceCiviliansArmed2021}. As a result, methods for clustering conflict events must incorporate an adjustable scale in order to engage with the full complexity of conflict \cite{zammit-mangionPointProcess2012}. 

We develop a systematic, data-driven, and scale-dependent procedure for extracting chains of causal events, or ``conflict avalanches,'' from observational data that could serve as the basic objects of conflict study (Figure~\ref{fig:method}). We construct conflict avalanches using a filter for statistical signatures of causality with a general measure of predictability, transfer entropy, which is a widely used measure for identifying hidden connections between system components \cite{schreiberMeasuringInformation2000,papanaReducingBias2011,staniekSymbolicTransfer2008}. Here, we use transfer entropy to build causal networks that connect local conflict events to one another. To do so, we start with the assumption that causal patterns can be detected from a reduced time series that only considers the appearance or absence of conflict, or binarization that ignores the magnitude of events. On one hand, this is a practical solution for handling general challenges in estimating statistics from a large state space. On the other hand, the power of the simplification is borne out in how we successfully identify related events, meaningful scales, and causal mechanisms hypothesized in the literature.

We discover a mesoscale at which conflict avalanches align with sociopolitical intuition (Figure~\ref{fig:mesoscale}), corresponding to separation scales on the order of a few days to months and tens of kilometers to hundreds. First, we recognize that the geographic scales recovered {\it a priori} range from 60\,km to 400\,km. Reassuringly, this is the typical distance between large neighboring towns and cities, which are important geographic pinning points for conflict (see Appendix~\ref{sec: africapolis}). The lower cutoff is much larger than any individual urban agglomeration and implies that conflict relations are not visible at microscopic precision. This could be because there is truly little statistical signal at such level of detail or from the limited resolution of the ACLED data set. Furthermore, the irregular shape of the mesoscale indicates that space and time scales are not independent of one another, or that looking at longer time scales is not equivalent to looking at longer spatial scales. Finally, we find that the avalanches in the mesoscale display a wide range of dynamical and spatial structures such as power law scaling. Such patterns indicate that conflict avalanches reflect long-range correlations between events. Thus, the mesoscale presents an interesting set of scales in which to focus on causal conflict patterns.

Perhaps surprisingly, conflict avalanches in the mesoscale group together events in a way that aligns with causal mechanisms proposed in the literature. We compare our conflict avalanches with heavily studied conflicts in Eastern Nigeria, Somalia, and Sierra Leone. In each of the cases, our method recovers recognizable clusters of events that align with actor groups and distinct time periods. Yet, we also find surprising connections when we draw regions of causal interaction. With Nigeria, we connect with non-negligible probability events that are identified as Fulani militia with the Boko Haram core, suggesting that these conflicts are related to one another. In Sierra Leone, we connect RUF government forces with events in neighboring Liberia, in line with allegations of troops crossing the border. For these examples of causal validation, we focus on relatively short scales at the bottom corner of the mesoscale, but at larger geographic scales we also discover causal influence regions that highlight regional conflict patterns \cite{marshallConflictTrends2006} (see Appendix~\ref{fig:interaction}). This suggests that beyond confirming known cases of causal relationships, our procedure can provide a way of predicting new ones to test, refine, or inspire new hypotheses.

As a step in this direction, we develop conflict zones of causal interaction (Figure~\ref{fig:regions}). The zones are convex hulls of the probability that a nearby event is grouped into a conflict avalanche with the seed events. This is a practical application of our work to a problem that has attracted much attention in the literature \cite{raleighIntroducingACLED2010,kikutaNewGeography2022}. It could, when coupled with expertise in the particular conflict zone of interest, enable better policy decisions 
and the impact of conflict on related phenomena such as poverty, segregation, and crime \cite{kikutaNewGeography2022}. Importantly, we introduce the flexibility of an adjustable spatiotemporal scale which can be crucial for detecting patterns that only emerge at certain levels of coarseness \cite{buhaugAccountingScale2005, oloughlinModelingData2014, scheffranClimateChange2012}. For example, the relationship between rainfall variability and conflicts is indiscernible when using a large temporal window such as an year as compared to monthly temporal window  \cite{coulibalyIdentifyingImpact2022}. Conflicts show correlation with climate-related disasters only when the period of analysis is less than three months \cite{schleussnerArmedconflictRisks2016}, and more generally the choice of scale is important for the connection between climate change and conflict \cite{abrahamsUnderstandingConnections2017}. Poverty and conflict is most meaningful at the subcountry level \cite{hegrePovertyCivil2009}. 
Our scale-adjustable scope provides a natural way of handling such variability to identify areas of potential interest or highlight unseen connections that deserve deeper investigation.

The need to bridge microscopic and macroscopic descriptions generalizes to other spreading social processes including unrest, migration, epidemics, and their relationship to conflict. As such, our approach has potential for wider use. For example, estimates of the eventual geographic extent of new activity can inform response planning. Our minimal approach may be especially helpful in this regard because it is difficult to gather detailed and accurate information in conflict regions \cite{muellerMonitoringWar2021}. As another example, event avalanches can feed into automated methods of pattern discovery by providing structured input for training machine learning algorithms \cite{cedermanPredictingArmed2017,hochProjectingArmed2021,geModellingArmed2022,zammit-mangionPointProcess2012}. Our event avalanches can provide groupings across a hierarchy of scales, which can be further enhanced by other properties that we have not considered here like sociodemographic factors. Thus, we address a fundamental need for a tool in both policy and quantitative analysis for handling multiscale social processes, and we pave the way to explore, rather than be limited by, the variability across scales.

\section*{Acknowledgements}
We thank the CSH Theory Group including Rudi Hanel, Jan Korbel, Tuan Pham, and Stefan Thurner for helpful discussions. E.D.L.~acknowledges funding from the Austrian Science Fund under grant number ESP 127-N. N.K.~acknowledges funding from the Austrian Federal Ministry for Climate Action, Environment, Energy, Mobility, Innovation and Technology, Funding Agreement number GZ 2021-0.664.668.

\section*{Data availability \& code}
We will publish a database containing the Voronoi cells and conflict avalanches on a web archive upon publication.

\clearpage
\setcounter{figure}{0}
\setcounter{equation}{0}
\renewcommand{\thefigure}{S\arabic{figure}}
\renewcommand{\thetable}{S\arabic{table}}
\renewcommand{\theequation}{S\arabic{equation}}

\clearpage
\appendix
\section{Dataset}\label{sec:data}

Our primary dataset is the Armed Conflict Location \& Event Data (ACLED) Project. This project collects real-time data on armed conflicts around the world with a focus on African states. The dataset is a collection of individual conflict events, defined as a single incidence of violence at a particular location and time involving at least two actors. In our analysis, we primarily focus on the date location of the conflict events, and we use other information including actor identities and event description for validation of the conflict avalanches.

There are event-based armed conflict datasets besides ACLED such as the Global Terrorism Database (GTD); the Integrated Crisis Early Warning System (ICEWS) dataset; the Phoenix event dataset; the Global Database of Events, Language, and Tone (GDELT); and the Uppsala Conflict Data Programme Georeferenced Event Dataset (UCDP GED) \cite{raleighSimilaritiesDifferences2019}.
We choose to use ACLED in our analysis because of two major reasons:
\begin{enumerate}
    \item Event-based armed conflict databases extract their information from various news reports from multiple sources. This can be done either manually by the help of human researchers and experts or can be scraped automatically from the news articles. Since we are focusing on Africa, we require a dataset which is curated manually by experts since most news articles in Africa are not in English and have local contexts that are difficult to scrape using automated systems. ACLED, GTD, and UCDP GED are the only three expert-curated datasets. The others are compiled using automated systems which tend to be heavily biased towards conflict events reported in English and French media \cite{raleighSimilaritiesDifferences2019} since currently automated systems are not designed to crawl through each and every local language media.
    \item ACLED covers all violent activities that occur both within and outside the context of a civil war, particularly violence against civilians, militia interactions, communal conflict and rioting. The other data sets do not. GTD focuses on ``terrorism'' only. 
    UCDP GED only records conflict events with at least one fatality. These definitions of armed conflicts are too restrictive. Therefore, ACLED is the most suitable dataset for our analysis among the available event-based datasets.
\end{enumerate}

In ACLED the conflict events are categorized into five major types. We mainly focus our analysis on conflict events that are categorized as ``Battles.'' According to the ACLED codebook, there are three different kinds of battles that we include in our Battles conflict
avalanches. As quoted from the codebook, these are defined as follows:

\begin{enumerate}
\item Battles - No change of territory: ``A battle between two violent armed groups where control of the contested location does not change. This is the correct event type if the government controls an area, fights with rebels and wins; if rebels control a location and maintain control after fighting with government forces; or if two militia groups are fighting. Battles take place between a range of actors.''

\item Battle - Non-state actor overtakes territory: ``A battle between two violent armed groups where
non-state actors win control of a location. If, after fighting with another force, a non-state group acquires control, or if two non-state groups fight and the group that did not begin with control acquires it, this is the correct event. There are few cases where opposition groups other than rebels acquire territory.''

\item Battle - Government regains territory: ``A battle between two violent armed groups where the government (or its affiliates) regains control of a location. This event type is used solely for government re-acquisition of control. A small number of events of this type include militias operating on behalf of the government to regain territory outside of areas of a government’s direct control (for example, proxy militias in Somalia which hold territory independently but are allied with the Federal Government).''

\end{enumerate}

\section{Algorithm for generating conflict avalanches}\label{Sec: Avalanche generation}
\begin{itemize}
\item \textbf{Choice of scale}


We devise a systematic method to navigate through the spatial and temporal scales of the problem independently of one another.

To set the spatial scale, we divide Africa into bins of approximately equal area. To do this, we generate a Voronoi tiling using a Poisson disc-sampling algorithm \cite{daviesPoissonDiscSampling}. The spatial scale is set by setting the average distance between the centers of two neighboring tiles to be approximately \textit{b} km. To set temporal scale, we divide the total number of days in the dataset into contiguous sequence of bins of duration \textit{a} days. One spatial bin $x$ and temporal bin $t$ together form a unique spatiotemporal combination to which any conflict event belongs (Figure~\ref{fig: bins}c). 

\item \textbf{Binarization}

We label a spatiotemporal bin with at least one conflict event as a $1$ and any without an event as a $0$. This procedure results in a binary time series of on/off values at each Voronoi cell (see movie in references \citenum{kushwahaVisualizationACLED2022} and \citenum{kushwahaVisualizationACLED2022a}).

\begin{figure*}[t]\centering
    \includegraphics[width=.7\linewidth]{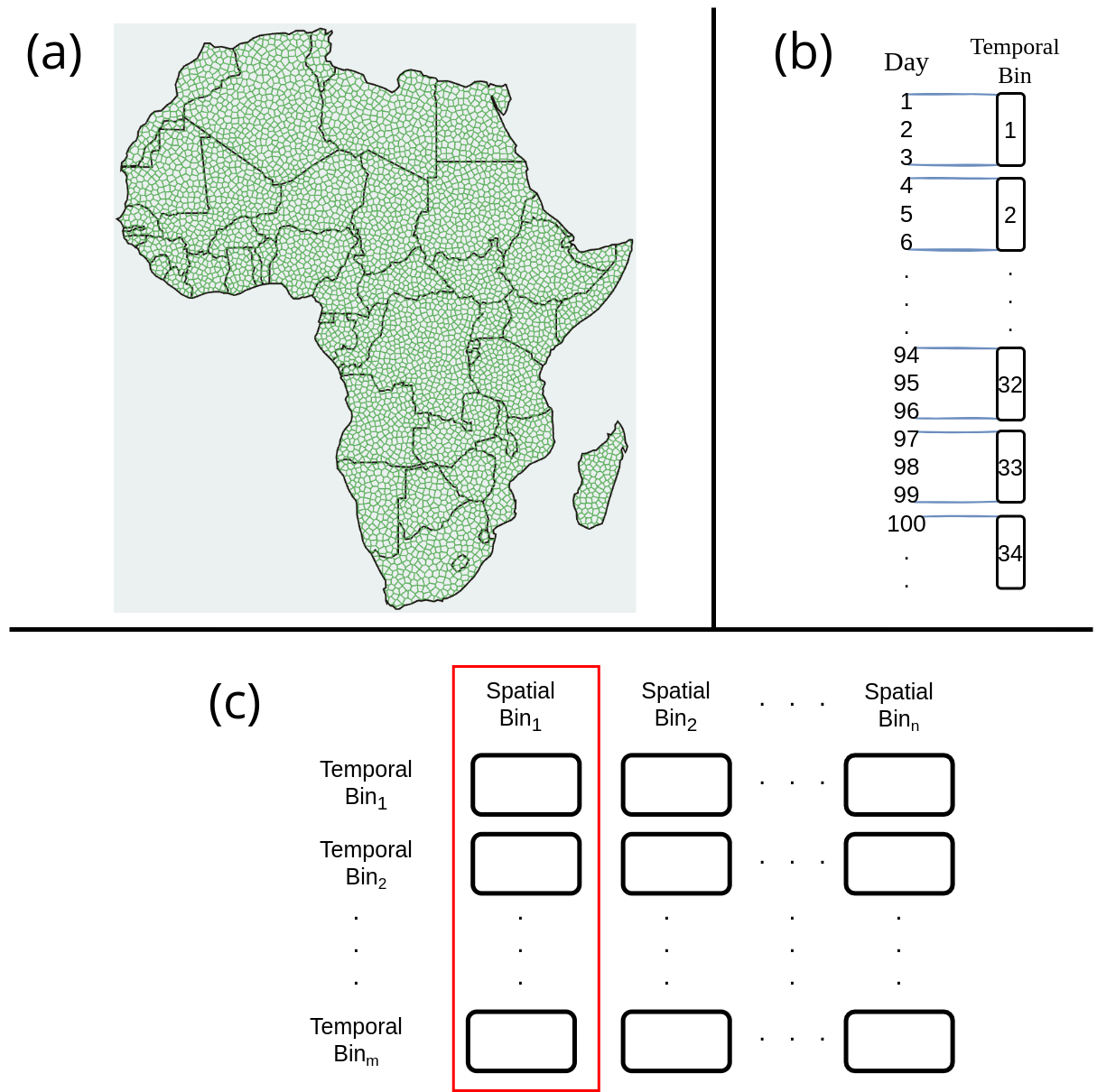}
    \caption{(a) Spatial bins. $b \approx 88$ km. (b) Temporal bins. In this example the temporal bins are of size $a=3$ days. (c) Spatiotemporal bins. Each box here is a spatiotemporal bin. Each spatiotemporal bin can have one of two values, one or zero. One represents presence of conflict and zero represents absence. Every spatiotemporal bin which has value one is called a "packet" of conflict event(s). Each column here is a binary time series of an individual spatial bin. For example, the spatiotemporal bins in the red box forms the time series for spatial bin number 1.}\label{fig: bins}
\end{figure*}

\item \textbf{Causal network}

We search for causal relationships between neighboring spatial bins from the statistics of on/off patterns with the transfer entropy (Eq~\ref{eq:te}). If the value of transfer entropy between a pair of spatial bins is significantly large with $p\leq 1/20$ with respect to a time-shuffled null model that erases temporal ordering of events at every Voronoi cell, we put a causal link between the bins. Since transfer entropy is asymmetric, we must calculate it in both directions for every pair of spatial bins. If the magnitude of transfer entropy is significant in both directions, we get a bi-directional causal link as we show in Figure~\ref{fig:network}. If the magnitude is significant only in one direction, we get a uni-directional causal link. We do this for all adjacent pairs of spatial bins to construct a causal network.

We search for a self-causal loop (an edge from a spatial bin at time $t$ to itself at time $t+1$) using the transfer entropy, which reduces to the mutual information in Eq~\ref{eq:I}.



\item \textbf{Clustering events}

We cluster together every pair of conflict events that satisfies one of the three following conditions:
\begin{itemize}
    \item The events occur in the same spatiotemporal bin $(t,x)$.
    \item The events are sequential in time bin and belong to the same tile with a self loop.
    \item The first event occurs at time $t$ and tile $x$, and the second occurs at time $t+1$ and another tile $x'$. There is a causal edge from $x$ to $x'$.
\end{itemize}
Once every pair has been clustered (some will remain alone), we have conflict avalanches. 
\end{itemize}

\section{Other ways of inferring causal network}
In principle, one could have constructed the ``causal'' network using other measures of temporal predictability such as Granger causality, time-delayed correlation, and time-delayed mutual information. We do not consider Granger causality because the variables we consider are not Gaussian, the assumption underlying that measure. While the latter measures do not explicitly distinguish the directionality of time because they are time-symmetric, we can still measure asymmetric information between sites by testing one site to be the past of the other and vice versa. 

\begin{figure}\centering
    \includegraphics[width=\linewidth]{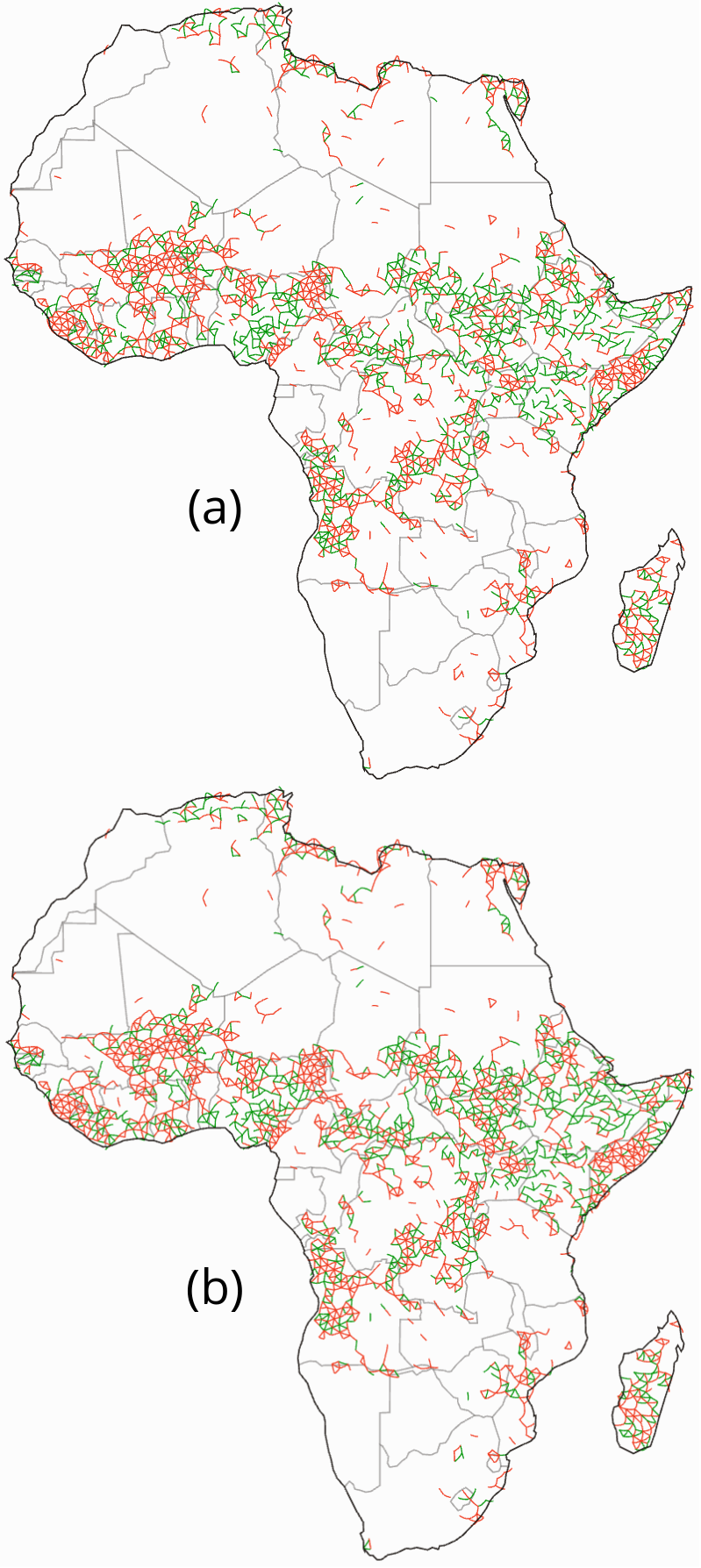}
    \caption{Conflict network inferred using (a) time-delayed pairwise correlation and (b) time delayed mutual information for $a=64$\,days, $b\approx88$\,km. We use 95\% bootstrapped confidence intervals to identify significant links between neighboring spatial bins. Directed nature of graph not shown. Edges shown in green have a causal edge in one direction only, red in both directions.}\label{fig:other networks}
\end{figure}

We show in Figure~\ref{fig:other networks} the resulting networks from the time-delayed pairwise correlation in panel a and with time-delayed mutual information in panel b. In the same way as with the transfer entropy calculation, we flag an edge as significant if it is of higher value than 95\% of bootstrapped time-shuffles. Unsurprisingly, the alternative measures return dense concentrations of links in similar areas as with the transfer entropy. In contrast, networks are denser both within the conflict hotspots and in more remote regions, suggesting that the transfer entropy provides a more discriminatory approach on which to build conflict avalanches.

\begin{figure}\centering
    \includegraphics[width=\linewidth]{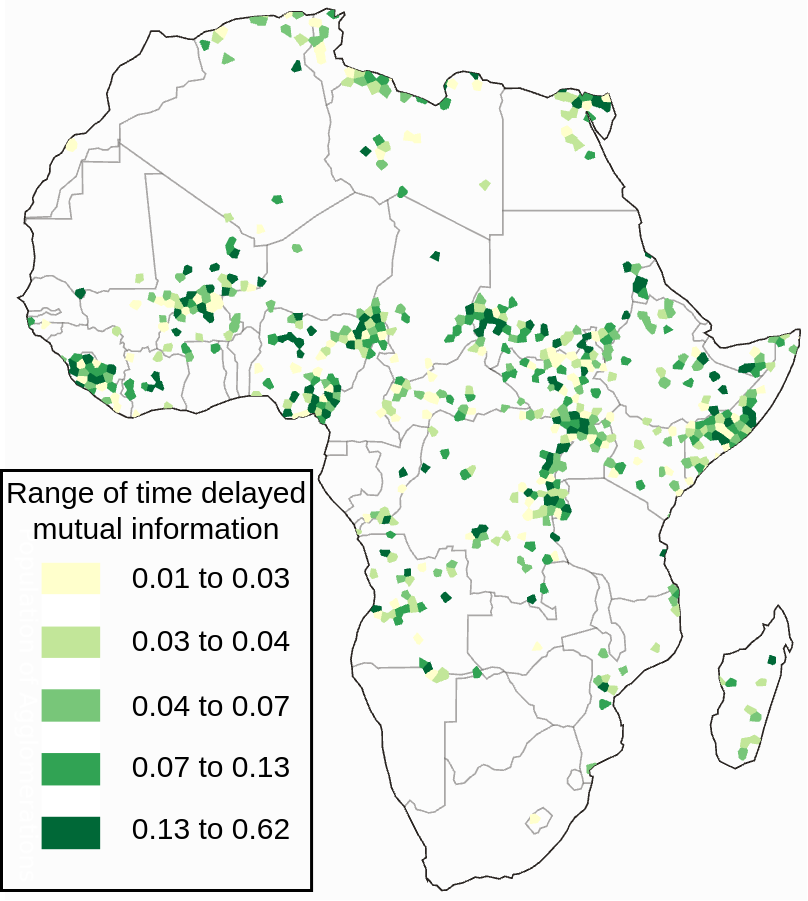}
    \caption{Time-delayed mutual information (Eq~\ref{eq:I}) at Voronoi cells ($\Dt=64$\,days, $\Dx\approx88$\,km).}\label{fig:self_loop}
\end{figure}

\begin{figure}\centering
    \includegraphics[width=\linewidth]{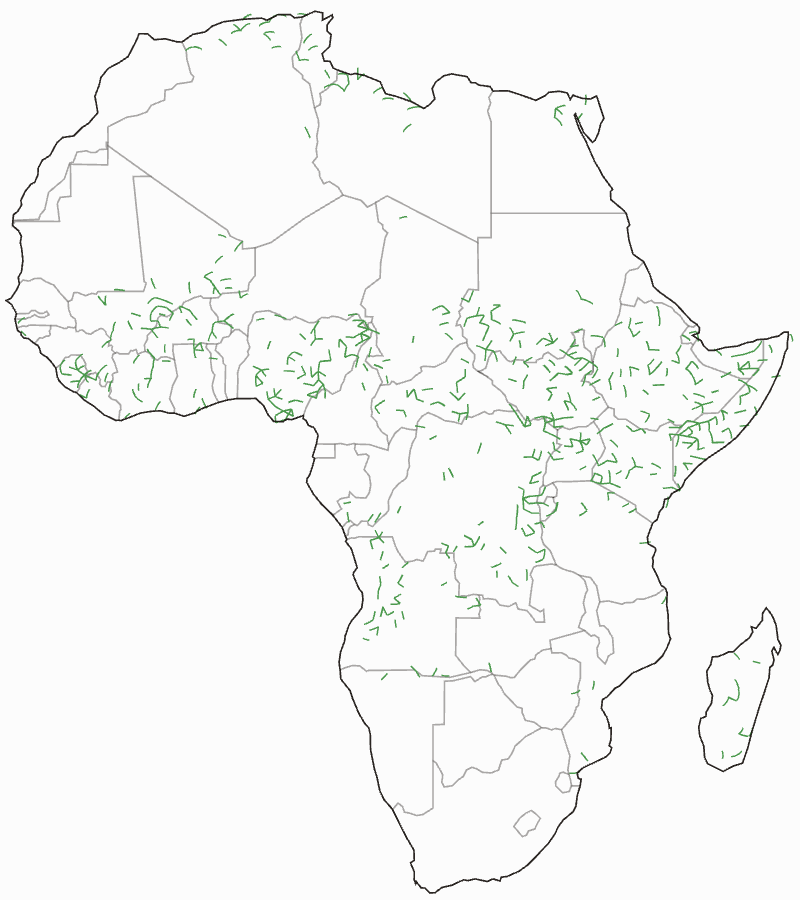}
    \caption{Causal network for space-shuffled null model. Temporal scale $\Dt=64$\,days and spatial scale $\Dx\approx88$\,km.}\label{fig: spatial null}
\end{figure}

\begin{figure}\centering
    \includegraphics[width=\linewidth]{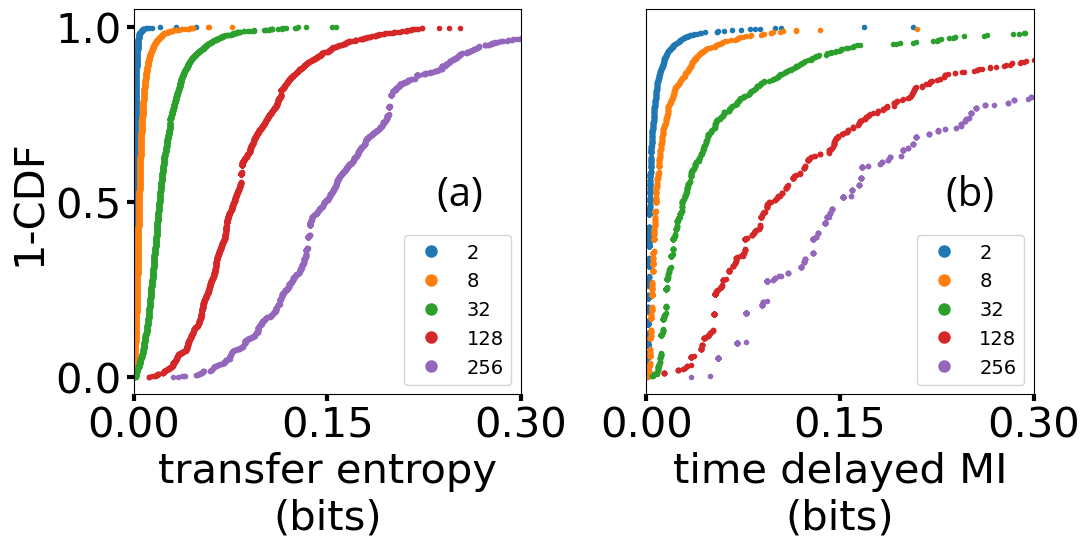}
    \caption{Complementary cumulative distribution function of (a) transfer entropy values of all valid links in a causal network and (b) mutual information between past and future of all self inciting spatial bins. Spatial scale $\Dx\approx88$\,km and different colors correspond to different temporal scale $a$.}\label{fig:te_distri}
\end{figure}

\section{Null models}
To test the procedure for extracting causal networks, we compare the causal network from data with two different randomized null models. 

The first null model is time-shuffled, where we randomly reshuffle the binary sequence of activity in every single Voronoi cell independently of one another. This removes all temporal correlations and thus renders the Voronoi cells independent of one another. When we perform the same causal network construction procedure on the time-shuffled network, we still identify a few edges as significant because the $p$-value threshold will yield false positives. However, the network is structurally different from what we show in Figure~\ref{fig:network}a because it fragments as we show in Figure~\ref{fig:network}b. 

The second null model is space-shuffled, where we reassign randomly the entire temporal sequence of activity from one Voronoi cell to another. We only do this for cells that have at least one conflict event recorded. Again, the causal network fragments as we show in Figure~\ref{fig: spatial null}. 

That the null models lose the interesting features such as the dense locales of intertwined conflict that otherwise appears in the data confirms that our results are not artifacts of our procedure.

\section{Modeling the distributions of conflict properties}\label{sec: other distributions}
We measure several properties that characterize conflict avalanches including the number of reports $R$, fatalities $F$, duration $T$, diameter $L$, and area measured by the number of cells $N$. 
When the properties are taken over the ensemble of avalanches, they are typically (though not always) consistent with power law tails in the mesoscale. In order to reach this conclusion, we compared the power law against three potential distributions including the lognormal and exponential. The probability distributions are defined in Table~\ref{si tab:distro}.

\begin{table}[]
\caption{Definitions of distributions used to fit conflict properties \cite{clausetPowerLawDistributionsEmpirical2009}.}\label{si tab:distro}
\begin{adjustbox}{width=\columnwidth,center}
\begin{tabular}{c|c|cc}
\multicolumn{1}{l|}{}       & \multirow{2}{*}{Distribution}          & \multicolumn{2}{c}{Equation $p(x) = c f(x)$}                                                                                                                                                             \\
\multicolumn{1}{l|}{}       &                                        & $f(x)$                                                                  & $c$                                                                                                                            \\ \hline
\multirow{6}{*}{\rotatebox{90}{Continuous}} & \multirow{2}{*}{Power law}             & \multirow{2}{*}{$x^{-\alpha}$}                                          & \multirow{2}{*}{$(\alpha -1) x_{\rm min}^{\alpha -1}$}                                                                             \\
                            &                                        &                                                                         &                                                                                                                                \\
                            & \multirow{2}{*}{Lognormal}             & \multirow{2}{*}{$\frac{1}{x} \exp(- \frac{(\ln(x)-\mu)^2}{2\sigma^2})$} & \multirow{2}{*}{$\sqrt \frac{2}{\pi \sigma^2} \left[ {\rm erfc} \left( \frac{\ln x_{\rm min}-\mu}{\sqrt 2 \sigma} \right) \right]^{-1}$} \\
                            &                                        &                                                                         &                                                                                                                                \\
                            & \multirow{2}{*}{Exponential}           & \multirow{2}{*}{$\exp(-\lambda x)$}                                     & \multirow{2}{*}{$\lambda \exp(\lambda x_{\rm min})$}                                                                               \\
                            &                                        &                                                                         &                                                                                                                                \\
                            \hline
\multirow{4}{*}{\rotatebox{90}{Discrete}}   & \multirow{2}{*}{Power law}             & \multirow{2}{*}{$x^{-\alpha}$}                                          & \multirow{2}{*}{$\frac{1}{\zeta \left( \alpha , x_{\rm min} \right)}$}                                                            \\
                            &                                        &                                                                         &                                                                                                                                \\
                            & \multirow{2}{*}{Exponential}           & \multirow{2}{*}{$\exp(-\lambda x)$}                                     & \multirow{2}{*}{$\left(1-\exp(-\lambda) \right)\exp(\lambda x_{\rm min})$}                                                        \\
                            &                                        &                                                                         &                                                                                                                                \\ \hline
\end{tabular}
\end{adjustbox}
\end{table}

\begin{figure}\centering
    \includegraphics[width=\linewidth]{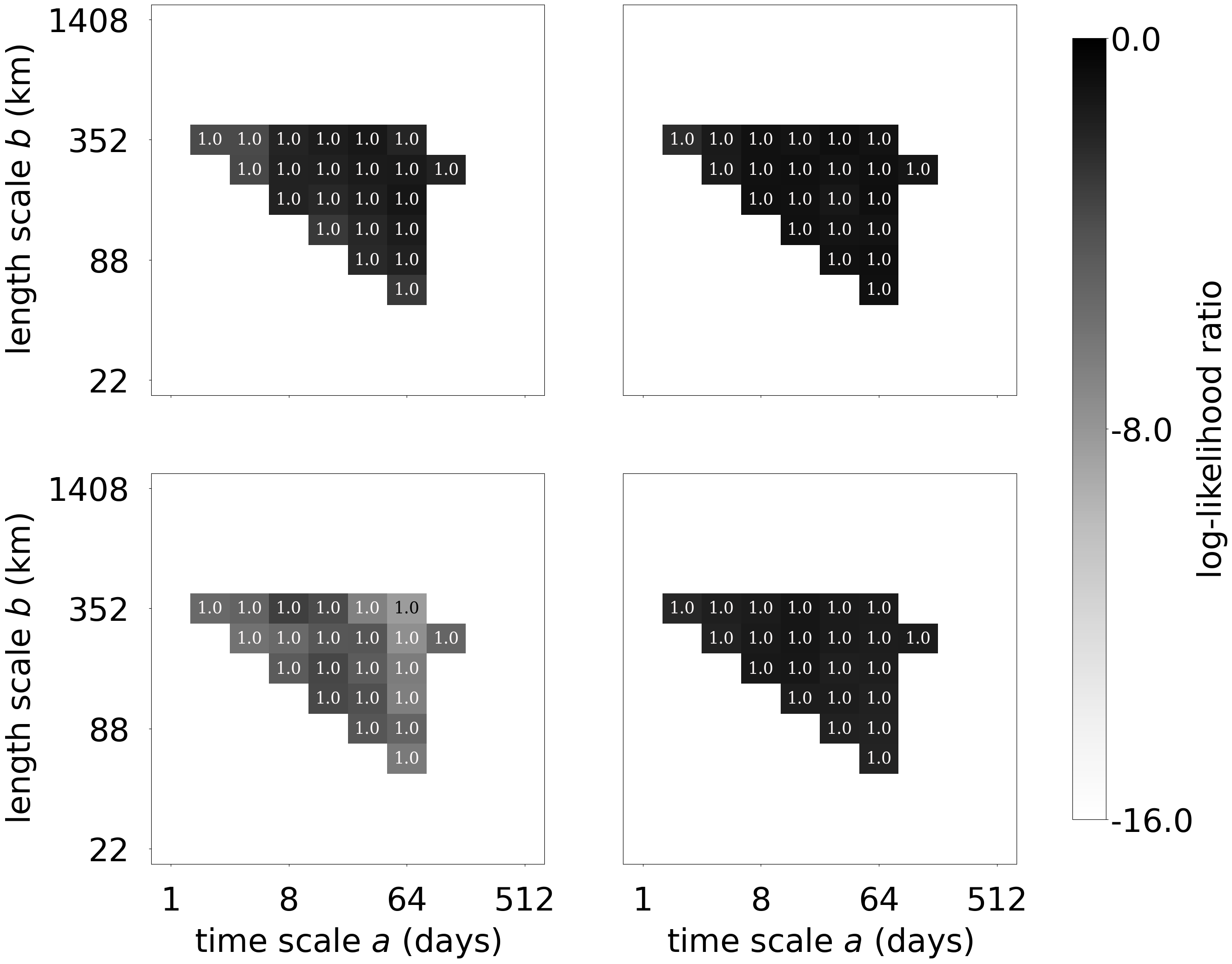}
    \caption{Comparison between power law and lognormal distribution for (a) fatalities, (b) reports, (c) sites, (d) duration. Each block is centered at a spatiotemporal scale (see Appendix~\ref{sec: interpolating} for more details) at which model distributions are compared by log-likelihood. Color bar shows the difference between the maximum log likelihood of lognormal to power law distribution averaged over 20 pseudorandom Voronoi tessellations. Number inside each spatiotemporal block shows the fraction of Voronoi realizations at which log-likelihood for power laws is greater.}\label{fig:ll_lognormal}
\end{figure}

\begin{figure}\centering
    \includegraphics[width=\linewidth]{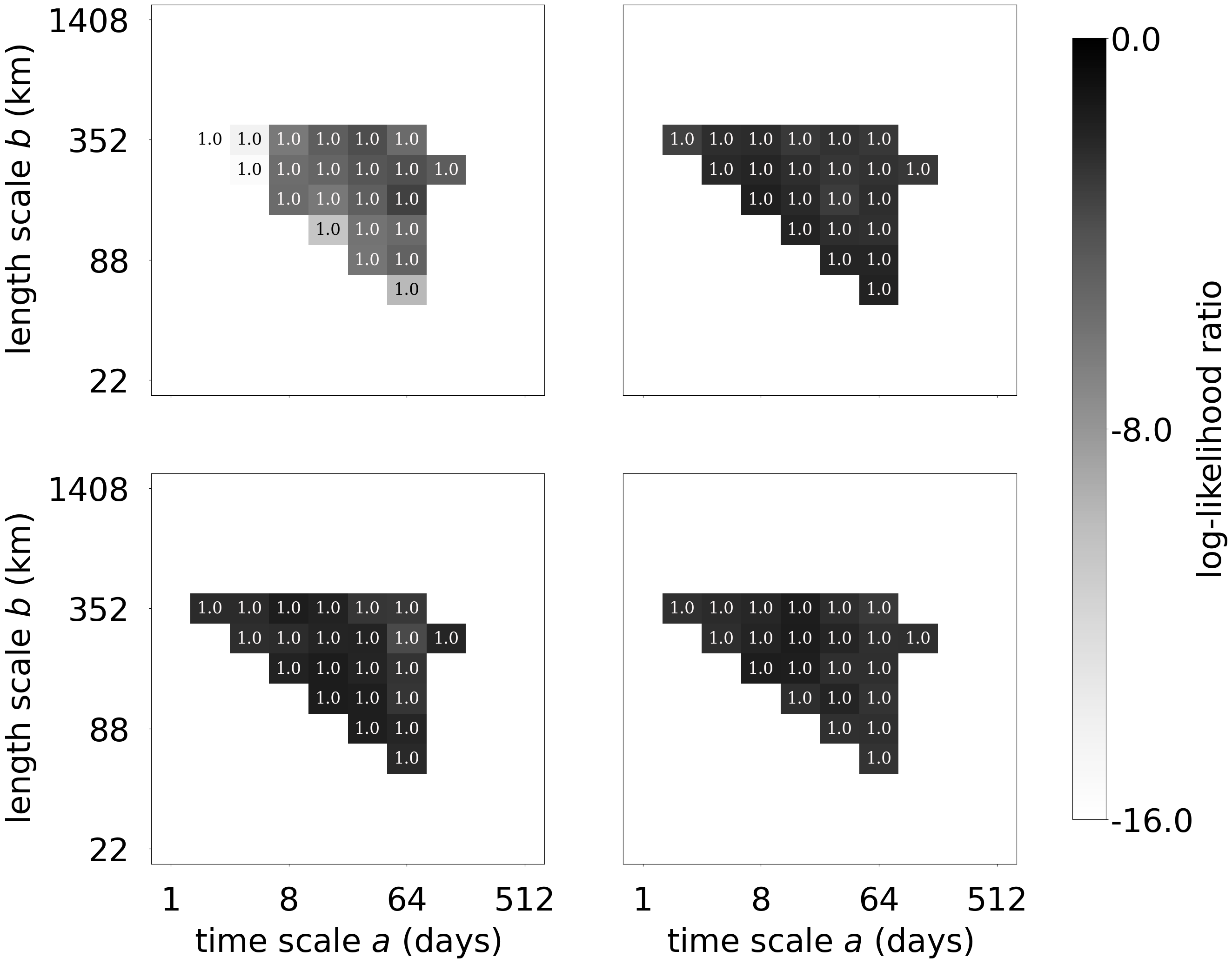}
    \caption{Comparison between power law and exponential distribution for (a) fatalities, (b) reports, (c) sites, (d) duration. Each block is centered at a spatiotemporal scale (see Appendix~\ref{sec: interpolating} for more details) at which model distributions are compared by log-likelihood. Color bar shows the difference between the maximum log likelihood of exponential to power law distribution averaged over 20 pseudorandom Voronoi tessellations. Number inside each spatiotemporal block shows the fraction of Voronoi realizations at which log-likelihood for power laws is greater.}\label{fig:ll_exponential}
\end{figure}

For each data point $x_{\rm k}$, we are able then to compute the likelihood of having observed it given the statistical model of the distribution. Looking over all the data points, we obtain the likelihood of the set of data
\begin{align}
	L(\{x_{\rm k}\}) &= \prod_{\rm k=1}^K p(x_{\rm k}),
\end{align}
where k is an index over the $K$ data points, and the probability according to the model is $p(x_{\rm k})$. Since the number of data points that we fit depend on the model --- for example, the power law comes with a lower cutoff --- we compute the typical log-likelihood for each data point that was fit by normalizing by the total number of data points considered $K$ such that we have $L/K$. According to the normalized log-likelihood on all the positively-valued data including the data points below the lower cutoff of the power law, the lognormal is superior to the power-model by a small amount (typically a factor of $\sim \sqrt{e}$). Upon comparing the fit to the tail of the distribution above the power law's lower cutoff in Figure~\ref{fig:ll_lognormal}, however, we find that the power-law is superior. This conforms with the additional statistical significance tests that we run, which show that the available data does not provide sufficient statistical resolution in many cases to rule out the power-law model. Both models are better fits to the data than an exponential as we show in Figure~\ref{fig:ll_exponential}. In this sense, we rely on the power-law model as a useful and approximate scaling hypothesis on which to relate the conflict properties to one another above some minimal scale.


\section{Population density}\label{sec: africapolis}
For gaining intuition about how the separation scales $\Dx$ and $\Dt$ relate to other social and geographic factors, we look at a map of population centers from the data set Africapolis. Africapolis considers population centers to be an ``urban agglomeration'' if the population exceed $10^4$ and there is no gap greater than 200 meters between built spaces \cite{oecdAfricaUrbanisation2020}. Population counts are extracted using census data and the built space is determined from satellite imagery. This provides a systematic and universal definition of a city, called an ``urban agglomeration,'' which does not depend on the vagaries of country records, datasets, and definitions. 

As we discuss in the main text, we use Africapolis to extract distances between the centers of agglomerations. We find that the mesoscale at which causal conflict patterns emerge corresponds to a geographic distance of 60\,km and 400\,km, which is the typical distance separating nearest agglomerations of above $10^4$ people and above $2{,}000{,}000$, respectively. These average distances, however, may differ between regions. For example, East Africa shows larger distances between large agglomerations. This suggests that when focusing on regions of Africa particular sections of the conflict mesoscale may be more revealing than others. Indeed, we find that when we compare the conflict regions obtained in Nigeria (which a relatively dense region) with those from Somalia (a relatively sparse regions), we find that the relevant separation scales to be smaller in the former than the latter. While population density is not a perfect correlate of the amount of armed conflict, we find that population density matters for extracting causal relationships between conflict activity.

\begin{figure}\centering
    \includegraphics[width=\linewidth]{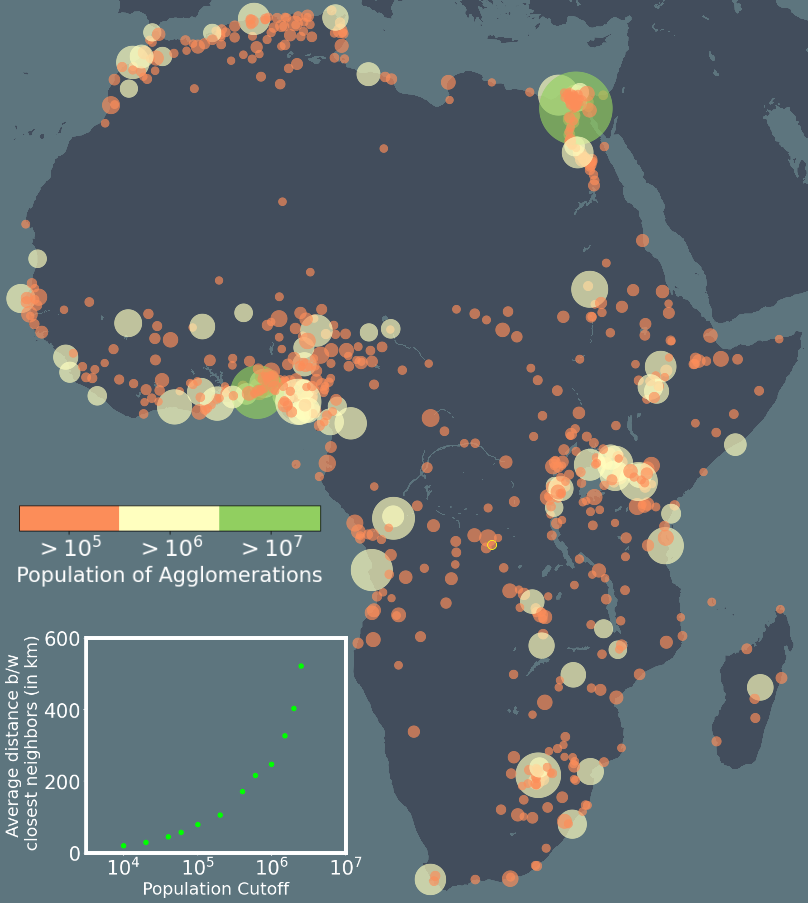}
    \caption{Urban agglomerations in Africa according to the Africapolis dataset. The radii of the circles are proportional to the agglomeration's population.}\label{fig:africapolis}
\end{figure}

\section{Actor similarity score}\label{sec:actor similarity}
\begin{figure}\centering
    \includegraphics[width=\linewidth]{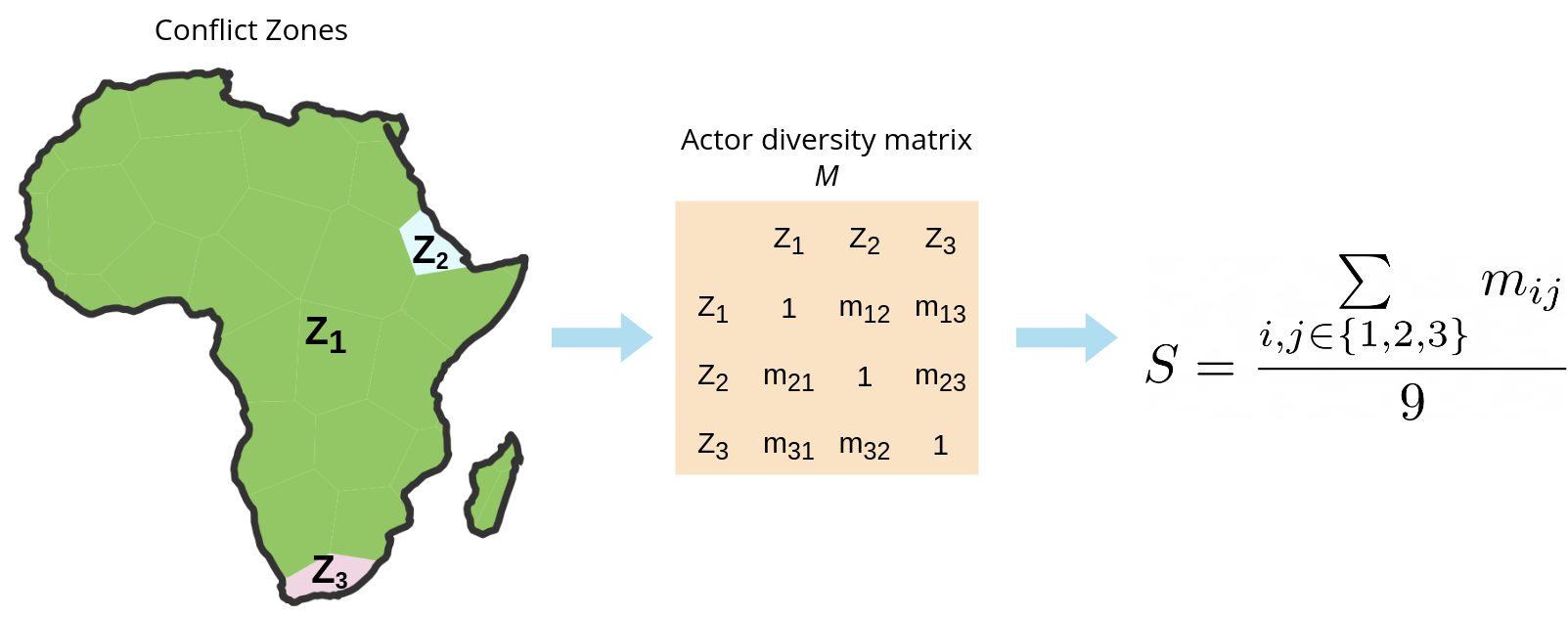}
    \caption{Visual representation for calculation of actor similarity $S$ at the scale $\Dt=8$\,days, $\Dx\approx1408$\,km. At the given scale we observe only three conflict zones in Africa. The equation to calculate the matrix elements $m_{\rm ij}$ is given in Eq~\ref{eqn:actor similarity}.}\label{fig:actor_similarity}
\end{figure}

To calculate the actor similarity score $S$, we first construct an actor similarity matrix $M$ comparing all pairs of conflict zones. Each element in the matrix accounts for the overlap between the sets of actors in each conflict zone weighted by the fraction of events in which they appear. The matrix $M$ is a square symmetric matrix with number of rows $=$ total number of conflict zones. Each element $m$ of this matrix is calculated as the overlap between a pair of conflict zones indexed i and j,
\begin{align}
	 m_{\rm ij} = \Theta_{Z_{\rm i}} \: . \: \Theta_{Z_{\rm j}}
  \label{eqn:actor similarity}
\end{align}
where, $\Theta_{Z_{\rm i}}$ is a vector with each entry as the fraction of events in zone $Z_{\rm i}$ involving each actor from the set of all observed actors. 
The actor similarity score $S$ is calculated by taking the mean of the actor similarity matrix $M$ at a given spatiotemporal scale (see Figure~\ref{fig:actor_similarity}).


\begin{figure}[t!]\centering
    \includegraphics[width=\linewidth]{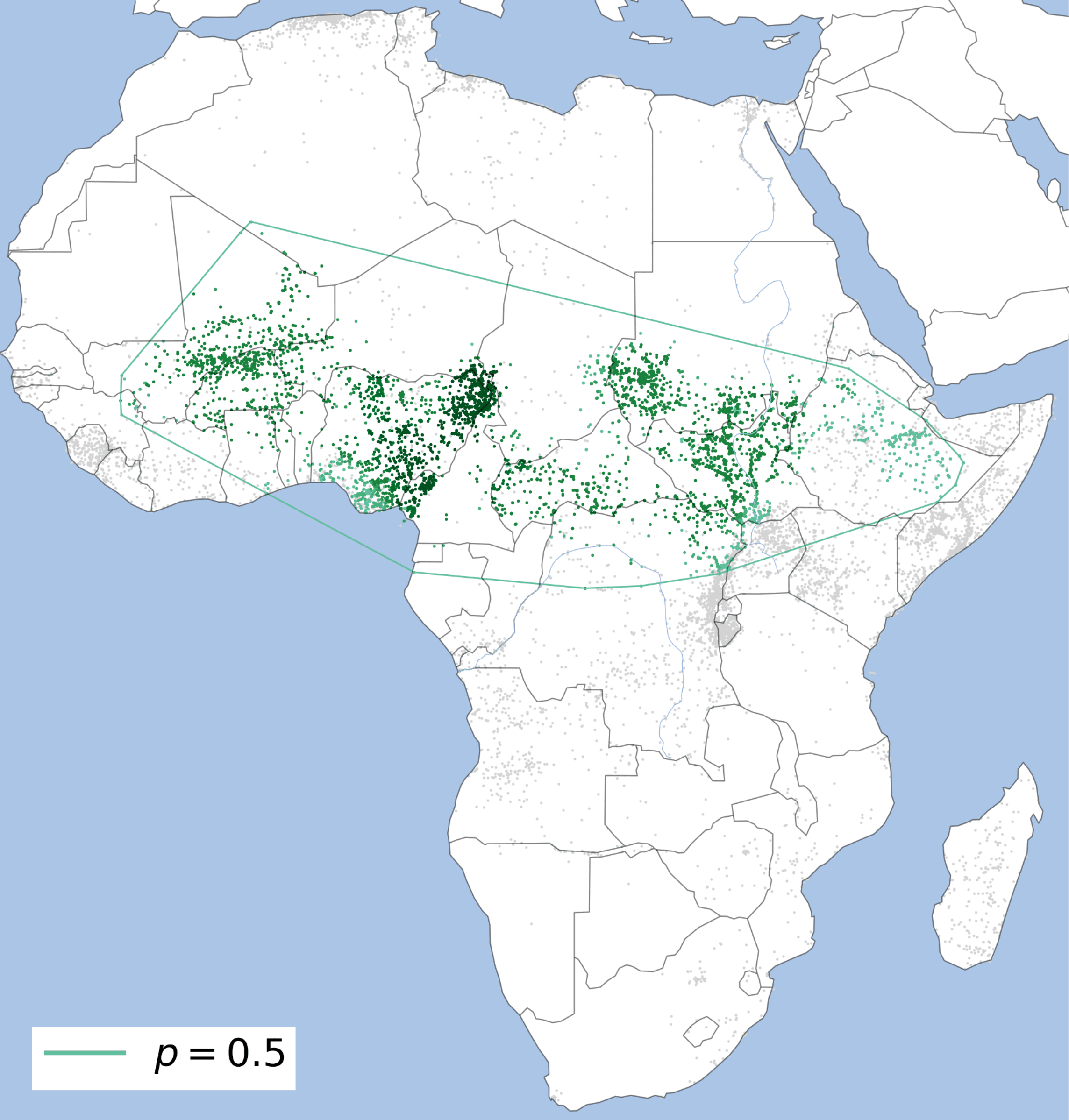}
    \caption{Conflict interaction zone for temporal scale $\Dt=64$\,days, spatial scale $\Dx\approx352$\,km, which is at the very top of the mesoscale in Figure~\ref{fig:mesoscale}.}
\label{fig:interaction}
\end{figure}

\section{Statistical causality and transfer entropy}\label{sec: Tranfer entropy}
In 1956, Wiener formulated causality in terms of predictability: ``For two simultaneously measured signals, if we can predict the first signal better by using the past information from the second one than by using the information without it, then we call the second signal causal to the first one'' \cite{wienerTheoryPrediction1956}. Later, Granger formulated it mathematically by introducing a statistical concept of causality based on evaluation of predictability which is now commonly known as ``Granger causality'' \cite{grangerInvestigatingCausal1969}. Transfer entropy is an information theoretic measure that generalizes some of the assumptions for Granger causality (GC). Namely, transfer entropy neither requires an explicit model (GC assumes a linear relationship between the predicted and predicting variables) nor assumes normality in their distributions. 

In short, transfer entropy is an information-theoretic quantity (typically based on Shannon entropy), which measures the flow of information between two or more time series which in turn is a measure for statistical causality. Recently transfer entropy has been used widely as a measure of statistical causality while dealing with complex systems in the form of time series \cite{vicenteTransferEntropy2011, wibralTransferEntropy2011, marschinskiAnalysingInformation2002}. 

Shannon entropy $H(X)$ is a measure which quantifies the amount of information that is needed to describe a system \cite{shannonMathematicalTheory1964}. It measures the average uncertainty of a system to be in a state $x$ out of all possible set of states $X$ such that $q(x)$ denotes the probability of the state $x$,

\begin{align}
    H(X) = -\sum\limits_{x \in X} q(x) \log q(x)
\end{align}

Mutual information is another information-theoretic quantity which measures the amount of information gained about one random variable $X$ through another random variable $Y$,

\begin{align}
    I[X;Y] &= \sum_{\substack{x \in X \\y \in Y}} q(x, y) \log\left( \frac{q(x,y)}{q(x)q(y)} \right)
\end{align}

Mutual information is a symmetric quantity, or $I[X;Y] = I[Y;X]$, and is a special case of the Kullback-Leibler divergence. 
In contrast, the transfer entropy measures the amount of directed (time-asymmetric) transfer of information between random processes. It quantifies the predictability of one variable by looking at the past value of another variable. This does not require making a model assumption as is the case with Granger entropy, but does require choosing the set of potentially informative past interactions (an assumption that also comes into Granger causality). The transfer entropy considers the ratio of the conditional distribution of one variable depending on the past samples of both processes versus the conditional distribution of that variable depending only on its own past values \cite{schreiberMeasuringInformationTransfer2000},

\begin{align}
	T[X;Y] &= \sum_{x_t,x_{t+1},y_t} q(x_t, x_{t+1}, y_{t}) \log \left(\frac{q(x_{t+1}|x_{t},y_{t})}{q(x_{t+1}|x_{t})}\right).
\end{align}

Transfer entropy is an asymmetric quantity, i.e $T[X;Y] \neq T[Y;X]$, and therefore unlike mutual information and correlation, transfer entropy can detect directional interactions which are a proxy for causal relationships. Importantly, the transfer entropy is by definition zero when two time-series are statistically independent, meaning that it can identify cases missing causal signatures.

\section{Interpolating mesoscale} \label{sec: interpolating}
In our study, we analyze a large range of spatial separation scales ($b\approx 22\,$km to $b\approx 1408\,$km) and temporal separation scales ($a=1\,$day to $a=512\,$days). To navigate through the scales, we analyze specific spatiotemporal scales and then interpolate the results in between. We calculate the spatial range at 13 marks separated by factors of $\sqrt{2}$. These are at about $b\approx 22,33,44,66,88,132,176,264,352,528,704,1056,1408\,$ km and the temporal range at 10 marks separated by factors of 2, $a=1,2,4,8,16,32,64,128,256,512\,$ days. Thus, we have $130$ combinations of spatiotemporal scales. In Figure~\ref{fig:mesoscale}b and d, we perform calculations for $130$ points in total and the rest of the space the values are computed using a bivariate spline approximation as implemented in scipy. Figure~\ref{fig:mesoscale}a is obtained by superimposing the contours from Figure~\ref{fig:mesoscale}b and d for $20$ realizations of the Voronoi tessellations. We obtain a smooth boundary with a bicubic interpolation.

\section{Scaling hypothesis} \label{sec: dyn_relations}

\begin{figure}\centering
    \includegraphics[width=\linewidth]{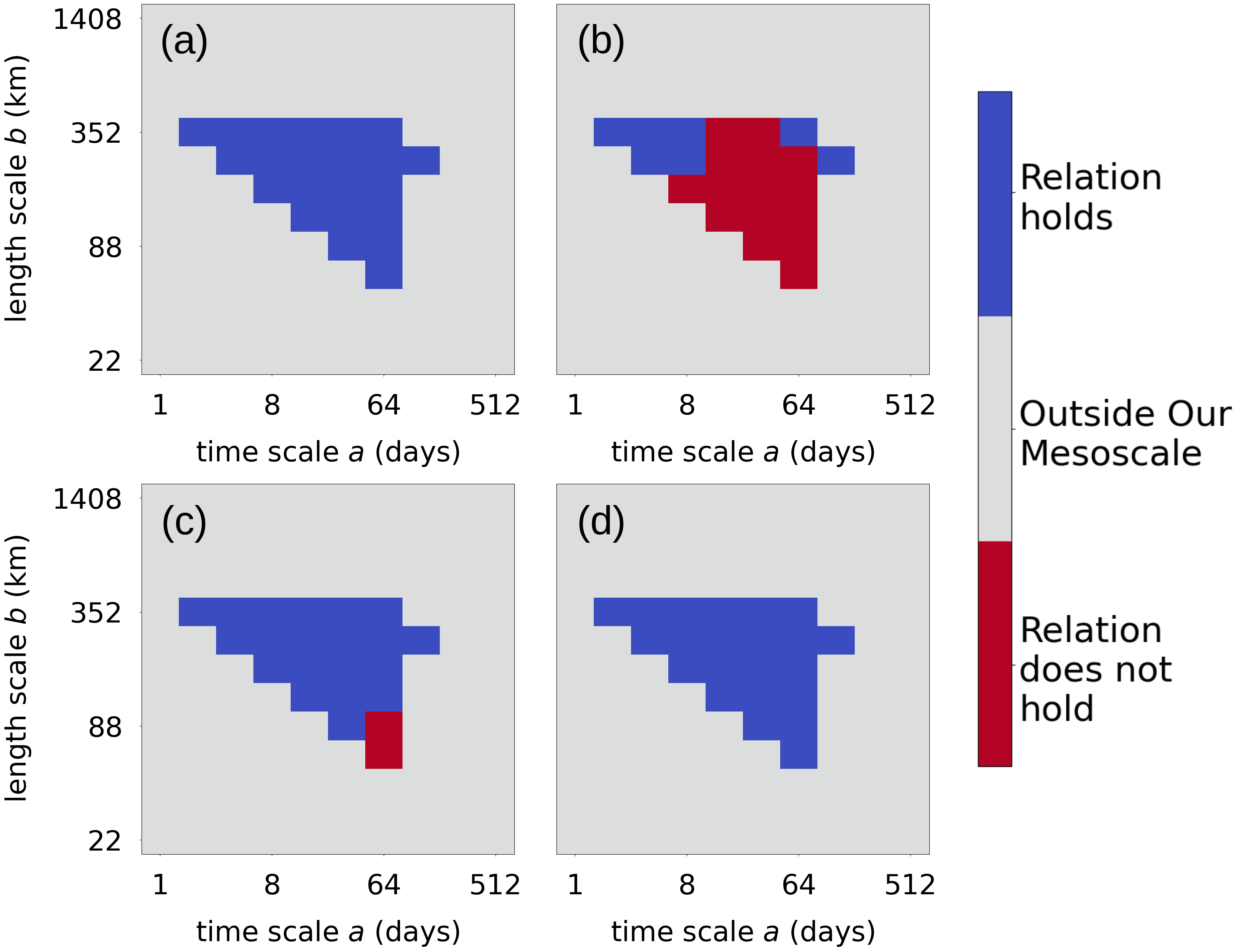}
    \caption{Significance of dynamical exponent relations in the mesoscale for (a) fatalities, (b) reports, (c), sites, and (d) diameter. Each block corresponds to a spatiotemporal scale (see Appendix~\ref{sec: interpolating} for more details). Validity of exponent relation is checked for 20 random realizations of Voronoi tessellations. If more than 90\% of these random realizations satisfy the exponent relation, we conclude that the exponent relation is significant (blue); otherwise, it is not (red).}\label{fig: dyn_relations}
\end{figure}

We check if the dynamical exponent relations relating each scaling variable with duration $T$ as discussed in the main text are satisfied across 20 random realizations of the Voronoi tessellations. If the relation is satisfied for a given spatiotemporal scale within the mesoscale more than 90$\%$ of time, we conclude that the exponent relation is significant. The result of this analysis is shown in Figure~\ref{fig: dyn_relations}.

\bibliography{merged}

\end{document}